\documentclass{ifacconf}

\usepackage{graphicx}      % include this line if your document contains figures
\usepackage{natbib}        % required for bibliography

\usepackage{graphicx}

\usepackage{amsmath}
\usepackage{amssymb}
\usepackage{amsfonts}
\usepackage{times}
\usepackage{mathptmx} % assumes new font selection scheme installed
\usepackage{datetime}

\DeclareMathAlphabet{\bit}{OML}{cmm}{b}{it}
\usepackage{datetime}
\usepackage{times} % assumes new font selection scheme installed
\usepackage{framed} % for pdf, bitmapped graphics files
\usepackage{graphicx} % for pdf, bitmapped graphics files
\usepackage{amsmath} % assumes amsmath package installed
\usepackage{amssymb}  % assumes amsmath package installed
\def\fH{\mathfrak{H}}

\def\<{\leqslant}           % nice less than or equal to sign
\def\>{\geqslant}           % nice larger than or equal to sign
\def\d{\partial}

\def\wt{\widetilde}

\def\Re{\mathrm{Re}}   % real part
\def\Im{\mathrm{Im}}   % imaginary part

\def\cH{\mathcal{H}}   % Hardy space
\def\mA{\mathbb{A}}    % space of real antisymmetric matrices
\def\mR{\mathbb{R}}    % real line
\def\Tr{\mathrm{Tr}}       % matrix trace
\def\rT{\mathrm{T}}        % matrix transpose
\def\bS{\mathbf{S}}

\def\bE{\mathbf{E}}    % expectation
\def\[[[{[\![\![}
\def\]]]{]\!]\!]}

\def\bra{\langle }
\def\ket{\rangle }

\def\re{\mathrm{e}}        % number e
\def\rd{\mathrm{d}}        % differential

\def\bA{\mathbf{A}}

\def\bJ{\mathbf{J}}

\def\br{\mathbf{r}}
\def\x{\times}
\def\ox{\otimes}
\def\op{\oplus}

\def\fF{{\mathfrak F}}

\def\fS{\mathfrak{S}}
\def\fP{\mathfrak{P}}

\def\cZ{\mathcal{Z}}

\def\cW{{\mathcal W}}

\def\cX{\mathcal{X}}

\def\cC{\mathcal{C}}

\def\sE{{\sf E}}

\def\cI{\mathcal{I}}

\def\cA{\mathcal{A}}
\def\cB{\mathcal{B}}

\def\cN{\mathcal{N}}

\def\cT{\mathcal{T}}

\def\mU{\mathbb{U}}

\def\ker{\mathrm{ker\,}}

\def\mS{\mathbb{S}}

\def\eps{\epsilon}

\def\ups{\upsilon}

\usepackage{datetime}

\begin{document}

\onecolumn
\pagestyle{plain}
\thispagestyle{empty}
\begin{center}
\title{\large A Homotopy Approach to Coherent Quantum LQG Control Synthesis Using Discounted Performance Criteria$^\star$}
\footnotetext[1]{This work is supported by
the Air Force Office of Scientific Research (AFOSR) under agreement number FA2386-16-1-4065
and the Australian Research Council under grant DP180101805.}

\author[IGV]{Igor G. Vladimirov$^*$, \qquad Ian R. Petersen}
\address[IGV]{Research School of Electrical, Energy and Materials Engineering, College of Engineering and Computer Science,
Australian National University, Canberra, Acton, ACT 2601,
Australia (e-mail: igor.g.vladimirov@gmail.com, i.r.petersen@gmail.com).}
\end{center}

\noindent                % Abstract of not more than 250 words.
-----------------------------------------------------------------------------------------------------------------------------------------

\noindent                % Abstract of not more than 250 words.
{\bf Abstract:} This paper is concerned with linear-quadratic-Gaussian (LQG) control  for a field-mediated feedback connection of a plant and a coherent (measurement-free) controller. Both the plant and the controller are multimode open quantum harmonic oscillators governed by linear quantum stochastic differential equations. The control objective is to make the closed-loop system internally stable and to minimize the infinite-horizon  quadratic cost involving the plant variables and the controller output subject to quantum physical realizability (PR) constraints. This coherent quantum LQG (CQLQG) control problem, which has been of active research interest for over ten years,  does not admit a solution in the form of separation principle and  independent Riccati equations known for its classical counterpart. We apply variational techniques
to a family of discounted CQLQG control problems parameterized by an effective time horizon. This gives rise to a homotopy algorithm, which is initialized with a PR (but not necessarily stabilizing) controller and aims at a locally optimal stabilizing controller for the original problem in the limit.


\noindent                % Abstract of not more than 250 words.
-----------------------------------------------------------------------------------------------------------------------------------------

%%%%%%%%%%%%%%%%%%%%%%%%%%%%%%%%%%%%%%%%%%%%%%%%%%%%%%%%%%%%%%%%%%%%%%%%%%%%%%%%%%%%%%%%%%%%%%%%%%%
\section{Introduction}
%%%%%%%%%%%%%%%%%%%%%%%%%%%%%%%%%%%%%%%%%%%%%%%%%%%%%%%%%%%%%%%%%%%%%%%%%%%%%%%%%%%%%%%%%%%%%%%%%%%

Control by interconnection aims to achieve certain dynamic properties of a plant through its interaction with other systems without digital signal processing. This control paradigm is particularly important in application to quantum systems whose variables are noncommuting operators on a Hilbert space governed by the laws of quantum mechanics and quantum probability \cite{H_2001,S_1994}. Using coherent quantum controllers with direct or field-mediated coupling \cite{JG_2010,ZJ_2011a} to quantum plants provides a measurement-free feedback architecture, which avoids back-action effects  and the loss of quantum information accompanying the process of measurement.

For open quantum harmonic oscillators (OQHOs), described by linear quantum stochastic differential equations (QSDEs) in the framework of the Hudson-Parthasarathy calculus \cite{HP_1984,P_1992}, the coherent quantum feedback leads to a fully quantum closed-loop system which is also organized as an OQHO \cite{NY_2017,P_2017}. Well-posedness (internal stability) and mean square performance criteria (quadratic cost minimization for the plant variables and controller output variables)  for such quantum systems  are similar in many respects to those in linear-quadratic-Gaussian (LQG) control theory for classical stochastic systems \cite{KS_1972}.
However, an essential distinction is the presence of physical realizability (PR) constraints \cite{JNP_2008},  which reflect the preservation of canonical commutation relations (CCRs) for quantum system variables and the parameterization of their dynamics in terms of energy and coupling matrices. These PR conditions make the coherent quantum LQG (CQLQG)  control setting  \cite{MP_2009,NJP_2009} a constrained covariance control problem which does not admit a solution in the form of the filtering-control separation principle and independent Riccati equations known in the classical case. Coherent quantum filtering problems \cite{MJ_2012,VP_2013b}, which are feedback-free versions of the control settings, also involve the PR constraints.
In the absence of classical separation structure, the CQLQG control problem can be solved numerically by using the gradient descent  \cite{SVP_2017} which employs Frechet differentiation of the mean square cost being minimized \cite{VP_2013a} and other variational and symplectic geometric techniques. This algorithm requires an internally stabilizing coherent quantum controller as an initial approximation, and the stabilization problem is also complicated by the PR constraints.

The present paper approaches the infinite-horizon CQLQG control problem as a limiting case of its discounted version, which employs (similarly to \cite{B_1965})  time averaging with an exponentially decaying weight specified by an effective time horizon (ETH). The discounted CQLQG control problem involves a relaxation of the internal stability constraint, which becomes inactive in the zero ETH limit and is essentially recovered as the ETH goes to infinity. We develop the first and second-order necessary conditions of optimality for the family of discounted problems, and use them for a zero-to-infinite horizon  homotopy algorithm, similar to \cite{MB_1985,VP_2018}, which is initialized with a (not necessarily stabilizing) coherent quantum  controller and aims at a locally optimal solution for the original problem in the limit.

The paper is organized as follows.
Section~\ref{sec:plant_cont} describes the class of quantum plants with coherent quantum feedback being considered.
Section~\ref{sec:PR} discusses the parameterization of the closed-loop system in terms of energy and coupling matrices.
Section~\ref{sec:CQLQGinf} describes the infinite-horizon coherent quantum LQG control problem and revisits the first-order conditions of optimality.
Section~\ref{sec:CQLQGT} specifies a family of CQLQG control problems with discounted mean square costs and provides first-order optimality conditions.
Section~\ref{sec:sec} discusses second-order conditions of optimality  and strongly locally optimal controllers for the discounted CQLQG control problems.
Section~\ref{sec:homo} describes a class of normal solutions for this family of problems and a homotopy differential equation.
Section~\ref{sec:conc} provides concluding remarks.

\section{Quantum plant and coherent quantum controller}
\label{sec:plant_cont}

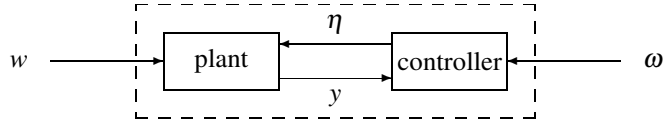
\begin{figure}[htbp]
\centering
\unitlength=1.5mm
\linethickness{0.2pt}
\begin{picture}(50.00,22.00)
    \put(7.5,10){\dashbox(35,10)[cc]{}}
    \put(10,12.5){\framebox(10,5)[cc]{plant}}
    \put(30,12.5){\framebox(10,5)[cc]{controller}}
    \put(0,15){\vector(1,0){10}}
    \put(50,15){\vector(-1,0){10}}
    \put(30,16.5){\vector(-1,0){10}}
    \put(20,13.5){\vector(1,0){10}}
    \put(-2,15){\makebox(0,0)[rc]{$w $}}
    \put(25,17.5){\makebox(0,0)[cb]{$\eta $}}
    \put(52,15){\makebox(0,0)[lc]{$\omega $}}
    \put(25,12.5){\makebox(0,0)[ct]{$y $}}
\end{picture}\vskip-12mm
\caption{The closed-loop system, resulting from a field-mediated interconnection of the quantum plant and the coherent quantum controller, is an open quantum system. It  interacts with the external input bosonic fields modelled by the quantum Wiener processes  $w $, $\omega $ which drive the QSDEs (\ref{x_y}), (\ref{xi_eta}). The plant-controller interaction is through the bosonic fields $y$, $\eta$ which are the plant output  and the controller output, respectively.
}
\label{fig:system}
\end{figure}

Consider a quantum plant and a coherent quantum controller which are organised as OQHOs interacting with each other in a measurement-free fashion through bosonic fields. In addition to the field-mediated interaction,
the plant and the controller also interact with external bosonic fields; see Fig.~\ref{fig:system}. %==============================================================================
%==============================================================================
The external fields are modelled by
quantum Wiener processes  $w_1, \ldots, w_{m_1}$ and $\omega_1, \ldots, \omega_{m_2}$ (with even $m_1$,  $m_2$) on symmetric Fock spaces \cite{HP_1984} $\fF_1$, $\fF_2$, respectively. These processes are assembled into the vectors
\begin{equation}
\label{womega}
  w:=
  (w_k)_{1\< k \< m_1},
    \qquad
  \omega:=
  (\omega_k)_{1\< k \< m_2},
    \qquad
    \cW
    :=
    {\begin{bmatrix}
        w\\
        \omega
    \end{bmatrix}},
\end{equation}
and have the Ito tables
\begin{equation}
\label{wwww}
    \rd w\rd w^{\rT} = \Omega_1 \rd t,
    \qquad
    \rd \omega\rd \omega^{\rT} = \Omega_2 \rd t,
    \qquad
    \rd \cW \rd \cW^{\rT} = \Omega \rd t   ,
\end{equation}
with the quantum Ito matrices $\Omega_1$, $\Omega_2$, $\Omega$ given by
\begin{align}
\label{JJJ}
    \Omega_k
    & := I_{m_k} + iJ_k,
    \qquad
    J_k:= \bJ \ox I_{m_k/2},
    \qquad
           \bJ
       :=
       {\begin{bmatrix}
         0 & 1\\
         -1 & 0
       \end{bmatrix}},
\\
\label{Om12}
    \Omega
    & :=
    {\begin{bmatrix}
        \Omega_1 & 0\\
        0 & \Omega_2
    \end{bmatrix}}
    =
    I_m + iJ,
    \qquad
    J:=
    {\begin{bmatrix}
        J_1 & 0\\
        0 & J_2
    \end{bmatrix}}    ,
    \qquad
    m:= m_1+m_2,
\end{align}
where $I_m$ is the identity matrix of order $m$, $\ox$ is the Kronecker product of matrices, and $\bJ$
spans the subspace of antisymmetric matrices of order 2. The augmented quantum Wiener process $\cW$ in (\ref{womega}) acts
on the tensor-product Fock space $\fF:= \fF_1\ox \fF_2$.
As OQHOs, the plant and the controller are endowed with initial Hilbert spaces $\fH_1$, $\fH_2$ and an even number $n$ of dynamic variables $x_1(t), \ldots, x_n(t)$ and $\xi_1(t),\ldots, \xi_n(t)$ which are time-varying self-adjoint operators on the space
\begin{equation}
\label{fH}
    \fH:= \fH_0\ox \fF,
\end{equation}
where $\fH_0:= \fH_1 \ox \fH_2$ is the initial plant-controller space. The plant and controller variables are assembled into the vectors
\begin{equation}
\label{xx}
    x
    :=
    (x_k)_{1\< k\< n},
    \qquad
    \xi
    :=
    (\xi_k)_{1\< k\< n},
    \qquad
    \cX
    :=
    {\begin{bmatrix}
        x\\
        \xi
    \end{bmatrix}}
\end{equation}
and satisfy the following CCRs with nonsingular matrices $\Theta_1, \Theta_2\in \mA_n$, $\Theta \in \mA_{2n}$ (with $\mA_n$ the subspace of real antisymmetric matrices of order $n$):
\begin{equation}
\label{Theta12}
    [\cX,\cX^{\rT}]
    =
    {\begin{bmatrix}
      [x,x^{\rT}] & [x,\xi^{\rT}]\\
      [\xi,x^{\rT}] & [\xi,\xi^{\rT}]
    \end{bmatrix}}
    =
    2i\Theta,
    \qquad
    \Theta
    :=
    {\begin{bmatrix}
      \Theta_1 & 0 \\
      0 & \Theta_2
    \end{bmatrix}}.
\end{equation}
The plant variables commute with the controller variables at every moment of time, that is, \begin{equation}
\label{xxicomm}
    [x,\xi^{\rT}]=0
\end{equation}
(in accordance with the block-diagonal structure of $\Theta$ in (\ref{Theta12})), since these operators  act initially (at time $t=0$)  on different spaces $\fH_1$, $\fH_2$, and the system-field evolution preserves the CCRs.
The output fields $y_1, \ldots, y_{p_1}$ and $\eta_1, \ldots, \eta_{p_2}$ of the plant and the controller (which mediate their interconnection) are time-varying self-adjoint operators on the space $\fH$ assembled into the vectors
$    y
    :=
    (y_k)_{1\< k\< p_1}$,
$
    \eta
    :=
    (\eta_k)_{1\< k\< p_2}
$.
The Heisenberg dynamics of the plant variables and the plant output are governed by linear QSDEs
\begin{equation}
\label{x_y}
    \rd x
    =
    A x \rd t  +  B \rd w  + E \rd \eta ,
    \qquad
    \rd y
    =
    C x \rd t  +  D \rd w ,
\end{equation}
where
\begin{equation}
\label{ABCDE}
    A\in \mR^{n\x n},
    \qquad
    B\in \mR^{n\x m_1},
    \qquad
    C\in \mR^{p_1\x n},
    \qquad
    D\in \mR^{p_1\x m_1},
    \qquad
    E\in \mR^{n\x p_2}
\end{equation}
are given matrices.
The feedthrough matrix $D$ is formed from conjugate pairs of rows of a permutation matrix of order $m_1$, so that $p_1$ is even and satisfies $p_1\< m_1$, with
\begin{equation}
\label{DDI}
    DD^{\rT} = I_{p_1},
\end{equation}
and hence, $D$ is of full row rank. The quantum Ito matrix $\wt{\Omega}_1$ of the plant output fields in (\ref{x_y}), defined by
$    \rd y \rd y^{\rT}
    =
    \wt{\Omega}_1 \rd t
$
and computed as     $\wt{\Omega}_1
    :=
    D\Omega_1 D^{\rT}
    = I_{p_1} + i\wt{J}_1$, has an orthogonal imaginary part
\begin{equation}
\label{tJ1}
    \wt{J}_1:= DJ_1D^{\rT}
\end{equation}
(that is, $\wt{J}_1^2= -I_{p_1}$ in view of $\wt{J}_1$ being antisymmetric).
The structure the matrices $A$, $B$, $C$, $E$ of the quantum plant  in (\ref{ABCDE}) is discussed in Section~\ref{sec:PR}.
The first QSDE in (\ref{x_y}) is driven by the external input field $w$ and the controller output $\eta$ which corresponds to the actuator signal  in  classical linear control theory  \cite{KS_1972}. Similarly, the second QSDE in (\ref{x_y}) for the plant output $y$ resembles the equations for noise-corrupted observations with a ``signal'' part
\begin{equation}
\label{z}
    z
    :=
     Cx.
\end{equation}
However, the quantum process $y$ is qualitatively different from the classical observations  because its entries are operator-valued, and the noncommutative quantum nature  of the output  fields $y_1, \ldots, y_{p_1}$ makes them inaccessible to simultaneous measurement. This noncommutativity is seen from the relation
$    [y(s), y(t)^{\rT}] = 2i \min(s,t) \wt{J}_1$ for all
    $s,t\>0$,
whose right-hand side does not vanish.
The dynamic variables and output fields of the coherent quantum controller satisfy the linear QSDEs
\begin{equation}
\label{xi_eta}
    \rd \xi
     =
    a\xi \rd t + b  \rd \omega  + e \rd y ,
    \qquad
    \rd \eta
     =
    c \xi  \rd t + d \rd \omega,
\end{equation}
with the matrices
\begin{equation}
\label{abcde}
    a  \in \mR^{n\x n},
    \qquad
    b \in \mR^{n\x m_2},
    \qquad
    c \in \mR^{p_2\x n},
    \qquad
    d \in \mR^{p_2\x m_2},
    \qquad
    e \in \mR^{n\x p_1}.
\end{equation}
Similarly to $D$ in (\ref{x_y}), (\ref{DDI}), the controller feedthrough matrix $d$ in (\ref{xi_eta}) is also of full row rank and consists of conjugate pairs of rows of a permutation matrix of order $m_2$, so that $p_2$ is even and satisfies $p_2\< m_2$, with
\begin{equation}
\label{ddI}
    dd^{\rT} = I_{p_2}.
\end{equation}
Accordingly, the quantum Ito matrix $\wt{\Omega}_2$ of the controller output fields in (\ref{xi_eta}), which is defined by
$    \rd \eta \rd \eta^{\rT} = \wt{\Omega}_2 \rd t$
and computed as
$    \wt{\Omega}_2
    :=
    d\Omega_2 d^{\rT}
    = I_{p_2} + i\wt{J}_2$, has an orthogonal imaginary part
\begin{equation}
\label{tJ2}
    \wt{J}_2:= dJ_2d^{\rT} \in \mA_{p_2},
\end{equation}
with $\wt{J}_2^2 = -I_{p_2}$.
In what follows, it is assumed that the matrix $d$ (which quantifies the ``amount'' of noise $\omega$ in the controller output $\eta$) is fixed, while the matrices $a$, $b$, $c$, $e$ in (\ref{abcde}) can be varied as specified in Section~\ref{sec:PR}. Similarly to (\ref{z}), the drift vector
\begin{equation}
\label{zeta}
    \zeta
     :=
    c \xi
\end{equation}
in (\ref{xi_eta})
plays the role of a ``signal'' part of the controller output $\eta$ as a quantum noise-corrupted actuator process.
The matrices $b $, $e $ in (\ref{xi_eta}) are the gain matrices of the controller with respect to the controller noise $\omega $ and the plant output $y $ in (\ref{x_y}).
The combined set of QSDEs
 (\ref{x_y}), (\ref{xi_eta}) governs
the fully quantum closed-loop system shown in Fig.~\ref{fig:system}.
In view of the analogy between the process $\zeta $ in (\ref{zeta}) and the actuator signal
 in the classical LQG approach,
the performance of the coherent quantum controller will be described
by using the infinite-horizon and discounted mean square cost functionals in terms of an $r$-dimensional process
\begin{equation}
\label{cZ}
    \cZ
     :=
     (\cZ_k)_{1\< k\< r}
     :=
    Fx   + G \zeta,
\end{equation}
where $
    F\in \mR^{r\x n}$,
$
    G\in \mR^{r\x p_2}
$ are given matrices. The entries of $\cZ$ are time-varying self-adjoint operators which are linear combinations of the plant variables and the controller output variables from (\ref{zeta}) whose relative importance is specified by the weighting matrices
$
    F$,
$
    G
$.
It is assumed that $G$ is of full column rank:
\begin{equation}
\label{Grank}
    r\> \mathrm{rank} G = p_2
\end{equation}
(or equivalently, $G^{\rT}G\succ 0$), which means that $G\zeta$ in (\ref{cZ}) is not reducible to a smaller number of linear combinations of $\zeta_1, \ldots, \zeta_{p_2}$ and \emph{all}  the entries of $\zeta$ will be penalized for large mean square values, similarly to the classical LQG control settings \cite{KS_1972}.  In other respects, the matrices $F$, $G$ in (\ref{cZ}) are not subjected to physical constraints, and their choice is dictated by the control design preferences.
The process $\cZ $ in  (\ref{cZ})  can be  expressed in terms of the  combined vector $\cX$ of the plant and controller variables in (\ref{xx}) governed by
 \begin{equation}
\label{closed_ZX}
    \rd \cX
      =
      \cA      \cX \rd t +   \cB       \rd \cW ,
      \qquad
    \cZ
     =
      \cC        \cX ,
\end{equation}
where the QSDE is driven by the augmented quantum Wiener process $\cW$ from (\ref{womega}) on the tensor-product Fock space $\fF$. The matrices  $\cA\in \mR^{2n\x 2n}$, $\cB\in \mR^{2n\x m}$, $\cC\in \mR^{r\x 2n}$ of the closed-loop system (\ref{closed_ZX}) are computed by combining the QSDEs (\ref{x_y}), (\ref{xi_eta}) with  (\ref{zeta}), (\ref{cZ}) as
\begin{equation}
\label{cABC}
    \cA
    :=
    {\begin{bmatrix}
        A & Ec\\
        eC & a
    \end{bmatrix}},
    \qquad
    \cB
    :=
    {\begin{bmatrix}
        B & Ed\\
        eD & b
    \end{bmatrix}},
    \qquad
    \cC
    :=
    {\begin{bmatrix}
        F & Gc
    \end{bmatrix}}.
\end{equation}
While the matrices $F$, $G$ in (\ref{cZ}) can be arbitrary,  the matrices $\cA$, $\cB$ of the QSDE in (\ref{closed_ZX}) are of specific structure \cite{JNP_2008} inherited by the fully quantum closed-loop system from the plant and controller.

\section{Energy operators  and physical realizability of the closed-loop system}
\label{sec:PR}

The plant-controller dynamics are specified by energy operators (the system Hamiltonian and the system-field coupling operators) which are quadratic and linear functions of the system variables.
More precisely,
the plant matrices $A$, $B$, $C$, $E$ in (\ref{x_y}) and the controller matrices $a$, $b$, $c$, $e$ in (\ref{xi_eta}) are given by
\begin{align}
\label{AB}
    A
    & =
    2\Theta_1(R_1 + M_1^{\rT}J_1 M_1 + L_1^{\rT}\wt{J}_2L_1),
    \qquad\
    B   = 2\Theta_1 M_1^{\rT},\\
\label{CE}
    C & =2DJ_1 M_1,
    \qquad
    E = 2\Theta_1 L_1^{\rT},\\
\label{ab}
    a
    & =
    2\Theta_2(R_2 + M_2^{\rT}J_2 M_2 + L_2^{\rT}\wt{J}_1L_2),
    \qquad
    b  = 2\Theta_2 M_2^{\rT},\\
\label{ce}
    c  & = 2dJ_2 M_2,
    \qquad\ \,
    e  = 2\Theta_2 L_2^{\rT},
\end{align}
where use is made of the matrices $\wt{J}_1$, $\wt{J}_2$ from (\ref{tJ1}), (\ref{tJ2}).
Here, $R_1\in \mS_n$ is the energy matrix of the plant (with $\mS_n$ the subspace of real symmetric matrices of order $n$), and   $M_1\in \mR^{m_1\x n}$, $L_1\in \mR^{p_2\x n}$ are the matrices of coupling of the plant with the external input field $w$ and the controller output $\eta$, respectively. Similarly, $R_2\in \mS_n$ is the energy matrix of the controller, and $M_2\in \mR^{m_2\x n}$, $L_2\in \mR^{p_1\x n}$ are the matrices of coupling of the controller with the external input field $\omega$ and the plant output $y$. The energy and coupling matrices parameterize the individual Hamiltonians
\begin{equation}
\label{H12}
    H_1:= \tfrac{1}{2} x^\rT R_1 x,
    \qquad
    H_2:= \tfrac{1}{2} \xi^\rT R_1 \xi
\end{equation}
and the vectors
${\small\begin{bmatrix}
  M_1\\ L_1
\end{bmatrix}} x$, ${\small\begin{bmatrix}
  M_2\\ L_2
\end{bmatrix}} \xi$ of coupling operators for the plant and controller.

%%%%%%%%%%%%%%%%%%%%%%%%%%%%%%%%%%%%%%%%%%%%%%%%%%%%%%%%%%%%%%%%%%%%%%%%%%%%%%%%%%%%%%%%%%%%%%%%%%%%%
\begin{lem}
\label{lem:closed_RM}
The energy matrix $R\in \mS_{2n}$ of the closed-loop system  (\ref{closed_ZX}) and the matrix $M \in \mR^{m\x 2n}$ of coupling between the system and the external input field $\cW$ in (\ref{womega}) can be computed as
\begin{equation}
\label{Rclos_Mclos}
    R  =
    {\begin{bmatrix}
      R_1                                       & \tfrac{1}{2}(L_1^{\rT}c +C^{\rT}L_2)\\
      \tfrac{1}{2}(c^{\rT}L_1+L_2^{\rT}C)   & R_2
    \end{bmatrix}},
    \qquad
    M  =
    {\begin{bmatrix}
      M_1 & D^{\rT}L_2 \\
      d^{\rT}L_1 & M_2
    \end{bmatrix}}.
\end{equation}
Here, $R_1$, $M_1$, $L_1$ and $R_2$, $M_2$, $L_2$  are the energy and coupling matrices of the plant  and the controller, respectively.
\hfill$\square$
\end{lem}
%%%%%%%%%%%%%%%%%%%%%%%%%%%%%%%%%%%%%%%%%%%%%%%%%%%%%%%%%%%%%%%%%%%%%%%%%%%%%%%%%%%%%%%%%%%%%%%%%%%%%
\begin{pf}
Substitution of the matrices $A$, $E$, $a$, $e$ from (\ref{AB})--(\ref{ce}) into the matrix $\cA$ in (\ref{cABC}) yields
\begin{equation}
\label{cA}
    \cA
    =
    2
    {\begin{bmatrix}
      \Theta_1(R_1 + M_1^{\rT}J_1M_1 + L_1^{\rT}\wt{J}_2L_1) & \Theta_1L_1^{\rT}c \\
      \Theta_2L_2^{\rT}C &  \Theta_2(R_2 + M_2^{\rT}J_2M_2 + L_2^{\rT}\wt{J}_1L_2)
    \end{bmatrix}}.
\end{equation}
On the other hand, similarly to the structure of the matrices $A$, $a$ in (\ref{AB}), (\ref{ab}),
\begin{equation}
\label{cAM}
    \cA = 2\Theta (R + M^{\rT}JM).
\end{equation}
Since the matrix $J$ in (\ref{Om12}) is antisymmetric, the closed-loop system  energy matrix $R\in \mS_{2n}$ is uniquely recovered from (\ref{cAM}) as
$
    R
    =
    \frac{1}{2}
    \bS(\Theta^{-1}\cA)
    =
    \bS
    \left(
    {\small\begin{bmatrix}
      R_1 + M_1^{\rT}J_1M_1 + L_1^{\rT}\wt{J}_2L_1           & L_1^{\rT}c \\
      L_2^{\rT}C                            &  R_2 + M_2^{\rT}J_2M_2 + L_2^{\rT}\wt{J}_1L_2
    \end{bmatrix}}
    \right)
$,
which leads to the first equality in (\ref{Rclos_Mclos}) in view of the symmetry of $R_1$, $R_2$ and antisymmetry of $J_1$, $J_2$. Here, $\bS(N):= \frac{1}{2}(N+N^\rT)$ is the symmetrizer of matrices, and use is also made of the block diagonal CCR matrix $\Theta$ of the plant and controller variables from (\ref{Theta12}).
By a similar reasoning, substitution of the matrices $B$, $E$, $b$, $e$ from (\ref{AB})--(\ref{ce}) into the matrix $\cB$ in (\ref{cABC}) yields
\begin{equation}
\label{cB}
    \cB
    =
    2
    {\begin{bmatrix}
      \Theta_1M_1^{\rT}& & \Theta_1L_1^{\rT}d\\
      \Theta_2L_2^{\rT}D & & \Theta_2M_2^{\rT}
    \end{bmatrix}}.
\end{equation}
On the other hand, similarly to the structure of the matrices $B$, $b$ in (\ref{AB}), (\ref{ab}), the coupling matrix $M$ of the closed-loop system is obtained from $\cB$ as $M = -\frac{1}{2}\cB^{\rT}\Theta^{-1}$, which leads to the second equality in (\ref{Rclos_Mclos}).\hfill$\blacksquare$
\end{pf}
%%%%%%%%%%%%%%%%%%%%%%%%%%%%%%%%%%%%%%%%%%%%%%%%%%%%%%%%%%%%%%%%%%%%%%%%%%%%%%%%%%%%%%%%%%%%%%%%%%%%%

In accordance with physical realizability (PR) conditions for OQHOs \cite{JNP_2008,SP_2012}, the closed-loop system matrices $\cA$, $\cB$ in (\ref{cA}), (\ref{cB}) satisfy
\begin{equation}
\label{PR}
    \cA \Theta + \Theta \cA^{\rT}
    +
    \cB J \cB^{\rT}
    =0.
\end{equation}
Here,
 $J=\Im \Omega$ from (\ref{Om12}) is the CCR matrix for the combined quantum Wiener process $\cW$ in  (\ref{womega}), (\ref{wwww}) in the sense that $[\rd \cW, \rd \cW^\rT] = 2iJ\rd t$.

Lemma~\ref{lem:closed_RM} (given above for completeness) can also be established by using the quantum feedback network formalism \cite{GJ_2009} which allows for calculation of global energy operators in terms of local parameters for arbitrary interconnections of quantum stochastic systems.
In view of (\ref{Rclos_Mclos}), the Hamiltonian of the closed-loop system is the following quadratic function of the plant and controller variables (\ref{xx}):
\begin{align}
\nonumber
  H
  & =
  \tfrac{1}{2}
  \cX^{\rT} R \cX\\
\nonumber
  & =
  \tfrac{1}{2}
  \big(
    x^{\rT} R_1 x + \xi^{\rT} R_2 \xi
  +
  x^{\rT}(L_1^{\rT}c +C^{\rT}L_2)\xi
  \big)\\
\label{Hclos}
  & =
  H_1 + H_2
  +
  \tfrac{1}{2}
  (\zeta^{\rT}L_1x +z^{\rT}L_2\xi).
\end{align}
Here, use is also made of the drift vectors $z$, $\zeta$ of the plant and controller outputs from (\ref{z}), (\ref{zeta}) and the commutativity (\ref{xxicomm}) between the plant and controller variables, which implies $[x,\zeta^\rT]=0$, $[z,\xi^\rT]=0$, $[z,\zeta^\rT]=0$. If the matrices $L_1$, $L_2$ (which quantify the field-mediated coupling between  the plant and the controller) vanish, then so also do the off-diagonal blocks of the energy and coupling matrices $R$, $M$ of the closed-loop system in (\ref{Rclos_Mclos}). In this case, the Hamiltonian $H$ in (\ref{Hclos}) reduces to the sum of the Hamiltonians (\ref{H12}).

Since $\det\Theta_2\ne 0$, the gain matrices $b$, $e$ of an arbitrary coherent quantum controller in (\ref{ab}), (\ref{ce}) are related by linear bijections to the corresponding coupling matrices $M_2$, $L_2$.
Moreover, the matrices  $a$, $c$ of such a controller are parameterized by the triple
\begin{equation}
\label{Pi}
    \Pi
    :=
    (R_2, b,e)\in \mS_n \x \mR^{n\x m_2}\x \mR^{n\x p_1}
    =:
    \mU
\end{equation}
as
\begin{equation}
\label{a_c}
    a
     =
    2\Theta_2 R_2
    -
    \tfrac{1}{2}
    (
        b J_2 b^{\rT}
        +
        e \wt{J}_1e^{\rT}
    )
    \Theta_2^{-1},
    \qquad
    c
    =
    -d
    J_2
    b^{\rT}
    \Theta_2^{-1}.
\end{equation}
Here, the Hamiltonian part $2\Theta_2 R_2\in \Theta_2 \mS_n$ of the matrix $a$ (in the sense of the symplectic structure specified by $\Theta_2$) depends linearly on the energy matrix $R_2$ of the controller, while its skew Hamiltonian part $- \frac{1}{2} ( b J_2 b^{\rT} + e \wt{J}_1e^{\rT} ) \Theta_2^{-1} \in \Theta_2 \mA_n$ is a quadratic function of $b$, $e$.
In view of the condition $\det\Theta_1\ne 0$, the matrices $A$, $C$
of the quantum plant  admit similar representations:
$    A
    =
    2\Theta_1 R_1 - \tfrac{1}{2}(B J_1 B^{\rT} + E \wt{J}_2E^{\rT})\Theta_1^{-1}$,
$
    C
    = -D J_1 B^{\rT}\Theta_1^{-1}$.
Regardless of a particular performance criterion, the relations (\ref{a_c}),
which couple the matrices $a$, $c$ to $b$, $e$, make the optimization of the coherent quantum controller (\ref{xi_eta}) qualitatively different from the classical optimal control problems. In particular, the second equality in (\ref{a_c}) shows that the coherent quantum controller needs an ``intake'' of the  external quantum noise $\omega$ (with $b\ne 0$)
in order to produce a useful output $\eta$ with a nonzero drift component $\zeta$ in (\ref{zeta}). Since $d$ is of full row rank due to (\ref{ddI}), and $J_2$, $\Theta_2$ are nonsingular, the linear map $\mR^{n\x m_2} \ni b\mapsto c \in \mR^{p_2\x n}$ in (\ref{a_c}) is surjective, so that $c$ can be assigned any value by an appropriate choice of $b$. In view of (\ref{a_c}), the closed-loop system matrix $\cA$ in (\ref{cABC}) depends on the controller triple $\Pi$ from (\ref{Pi}) in a quadratic fashion:
\begin{equation}
\label{cAPi}
    \cA
    =
    {\begin{bmatrix}
      A & &-E d
    J_2
    b^{\rT}
    \Theta_2^{-1} \\eC &  &
          2\Theta_2 R_2
    -
    \tfrac{1}{2}
    (
        b J_2 b^{\rT}
        +
        e \wt{J}_1e^{\rT}
    )
    \Theta_2^{-1}
    \end{bmatrix}}.
\end{equation}
This dependence complicates the problem of finding an internally stabilizing coherent quantum controller (which makes the matrix $\cA$ Hurwitz). Such system stabilization under PR constraints is part of mean square optimal coherent quantum control problems.

\section{Infinite-horizon coherent quantum LQG control problem}
\label{sec:CQLQGinf}

Similarly to the LQG paradigm for classical stochastic systems \cite{KS_1972}, the averaged behaviour of OQHOs can be quantified in terms of mean square performance criteria.
For the closed-loop system in (\ref{closed_ZX}),  an infinite-horizon quadratic cost is provided by
\begin{equation}
\label{V}
    V
    :=
    \tfrac{1}{2}
    \lim_{T\to +\infty} \Big(\tfrac{1}{T}\int_0^T \bE (\cZ(t)^{\rT} \cZ(t)) \rd t\Big),
\end{equation}
Here, $\bE \varphi := \Tr(\rho \varphi)$ is the quantum expectation over an underlying density operator $\rho:= \rho_0\ox \ups$ on the system-field space $\fH$ in (\ref{fH}), with $\rho_0$ the initial  quantum state of the plant and controller on $\fH_0$, and $\ups$ is the field state on the Fock space $\fF$.
Also,
\begin{equation}
\label{cZcZ}
    \cZ^{\rT} \cZ = \sum_{k=1}^r \cZ_k^2 = \cX^{\rT} \cC^{\rT}\cC \cX
\end{equation}
is a time-varying positive semi-definite quantum variable
whose expectation yields a nonnegative-valued integrand in (\ref{V}). In view of
the orthogonality of the subspaces $\mS_n$, $\mA_n$ in the sense of the Frobenius inner product $\bra \cdot, \cdot\ket$ of matrices \cite{HJ_2007}, (\ref{closed_ZX}), (\ref{V}), (\ref{cZcZ}) imply that \cite{NJP_2009}
\begin{equation}
\label{VP}
    V =
    \tfrac{1}{2}\bra \cC^{\rT}\cC, P\ket,
    \qquad
    P:=
    \lim_{T\to +\infty}
    \Big(
        \tfrac{1}{T}
        \int_0^T
        \Xi(t)
        \rd t
    \Big),
\end{equation}
where
\begin{equation}
\label{Xi}
  \Xi(t):= \Re \bE (\cX(t)\cX(t)^\rT),
  \qquad
  t\>0,
\end{equation}
 with $\Im \bE (\cX(t)\cX(t)^\rT) = \Theta$ the CCR matrix from (\ref{Theta12}).
In what follows, the state $\ups$ of the input field $\cW$ is assumed to be the vacuum state \cite{P_1992}. Then, for any stabilizing controller and any initial system state $\rho_0$ with finite second moments, that is,
\begin{equation}
\label{cX0good}
    \bE(\cX(0)^\rT\cX(0))
    <
    +\infty,
\end{equation}
the function $\Xi$ in (\ref{Xi}) satisfies the Lyapunov ODE
\begin{equation}
\label{LODE}
    \dot{\Xi} = \cA \Xi +\Xi \cA^\rT + \cB\cB^\rT
\end{equation}
(with $\dot{(\ )}$ the time derivative) and takes the form
\begin{equation}
\label{Sigma1}
  \Xi(t)
  =
  \re^{t\cA} \Sigma\re^{t\cA^\rT}
  +
  \int_0^t
  \re^{\tau\cA}
  \cB \cB^\rT
  \re^{\tau\cA^\rT}
  \rd \tau,
\end{equation}
where
\begin{equation}
\label{P0}
  \Sigma:= \Xi(0)=\Re \bE (\cX(0)\cX(0)^\rT).
\end{equation}
In view of (\ref{Sigma1}) and since $\cA$ is Hurwitz, the matrix $P$ in (\ref{VP}) is the unique solution
\begin{equation}
\label{Plim}
    P
    =
  \lim_{t\to +\infty}
  \Xi(t)
  =
      \int_{\mR_+}
    \re^{t\cA} \cB\cB^{\rT}\re^{t\cA^{\rT}}
    \rd t
\end{equation}
of the algebraic Lyapunov equation (ALE)
\begin{equation}
\label{PALE}
  \cA P+ P\cA^\rT + \cB\cB^\rT = 0
\end{equation}
and coincides with the infinite-horizon controllability Gramian \cite{KS_1972} of the pair $(\cA, \cB)$. In this case, the closed-loop system variables have a unique invariant Gaussian quantum state \cite{KRP_2010} with zero mean vector and the quantum covariance matrix
\begin{equation}
\label{PTheta}
    S:=
    P + i\Theta
    =
    \int_{\mR_+}
    \re^{t\cA}
    \cB\Omega \cB^{\rT}
    \re^{t\cA^{\rT}}
    \rd t
    \succcurlyeq 0,
\end{equation}
whose positive semi-definiteness is stronger than $P\succcurlyeq 0$ and reflects the Heisenberg uncertainty relation \cite{H_2001}. The equality in (\ref{PTheta}) is obtained by combining (\ref{PALE}) with the PR property (\ref{PR}) and using the quantum Ito matrix $\Omega \succcurlyeq 0$ from  (\ref{Om12}), which leads to the ALE
$    \cA S +S\cA^\rT   + \cB \Omega \cB^\rT = 0$.
The strict inequality $S \succ 0$  in (\ref{PTheta}) holds if and only if the pair $(\cA,\cB\sqrt{\Omega})$ is controllable.

In view of (\ref{VP}), (\ref{PALE}), the mean square cost functional $V$ coincides (up to the factor of  $\frac{1}{2}$) with the squared $\cH_2$-norm of the transfer function for an auxiliary  classical linear system with the state-space realization triple $(\cA,\cB,\cC)$ and also admits the following representations:
\begin{equation}
\label{VQH}
    V
    =
    \tfrac{1}{2}\bra Q, \cB\cB^{\rT}\ket
    =
    -\bra \cA, \Gamma\ket.
\end{equation}
Here, associated with any stabilizing controller is the infinite-horizon observability Gramian
\cite{KS_1972} of the pair $(\cA,\cC)$:
\begin{equation}
\label{Qlim}
    Q:= \int_{\mR_+}
    \re^{t\cA^{\rT}}
    \cC^{\rT}\cC
    \re^{t\cA}
    \rd t,
\end{equation}
which is the unique solution of the ALE
\begin{equation}
\label{QALE}
  \cA^{\rT}Q+ Q\cA + \cC^{\rT}\cC = 0.
\end{equation}
Also,
\begin{equation}
\label{Hank}
  \Gamma:= QP
\end{equation}
is a diagonalizable matrix whose eigenvalues are the squared Hankel singular
values \cite{KS_1972} for the state-space realization triple $(A,B,\cC)$.  The matrix $\Gamma$ will be referred to as the \emph{Hankelian} of the closed-loop system.

Now, the coherent quantum LQG (CQLQG) control problem \cite{NJP_2009} is formulated as the minimization
\begin{equation}
\label{CQLQG}
  V\to \inf,
  \qquad
  \Pi \in \fP
\end{equation}
of the mean square cost (\ref{V}) (computed in (\ref{VP}) or (\ref{VQH})) over the matrix triples (\ref{Pi}) from the set
\begin{equation}
\label{fP}
  \fP
  :=
  \{
    \Pi \in \mU:\
    \cA\ {\rm in}\ (\ref{cAPi})\ {\rm is\ Hurwitz}
  \},
\end{equation}
which reflects the internal stability requirement. With the set $\mU$ in (\ref{Pi}) being regarded as a Hilbert space with the direct-sum inner product $\bra (r_1,b_1, e_1), (r_2,b_2, e_2)\ket_\mU:= \bra r_1, r_2\ket + \bra b_1, b_2\ket+\bra e_1, e_2\ket$, the functional  $V$ is Frechet differentiable on the open subset $\fP\subset \mU$. The  $\mU$-valued Frechet derivative
\begin{equation}
\label{dVdPi}
    \d_\Pi V = (\d_{R_2} V, \d_b V, \d_e V)
\end{equation}
with respect to the controller parameters is provided below \cite[Lemma~3, Theorem~1]{VP_2013a}.

%%%%%%%%%%%%%%%%%%%%%%%%%%%%%%%%%%%%%%%%%%%%%%%%%%%%%%%%%%%%%%%%%%%%%%%%%%%%%%%%%%%%%%%%%%%%%%%%%%%%%%%%%%%%%%%
\begin{thm}
\label{th:dVdRbe}
Suppose the closed-loop system (\ref{closed_ZX}) is driven by vacuum fields, and
the initial plant and controller variables in (\ref{xx}) satisfy (\ref{cX0good}). Then, for any stabilizing coherent quantum controller, described by (\ref{xi_eta}), (\ref{a_c}), the partial Frechet derivatives of the mean square cost $V$ in (\ref{V}) with respect to the matrices $R_2$, $b$, $e$ from (\ref{fP}) can be computed as
\begin{align}
\label{dVdR}
    \d_{R_2} V
    &=
    -2\bS(\Theta_2 \Gamma_{22}),\\
\label{dVdb}
    \d_b V
    & =
        Q_{21}Ed + Q_{22} b-\bA(\Gamma_{22}\Theta_2^{-1})bJ_2
    -
    \Theta_2^{-1}
    (\Gamma_{12}^{\rT} E + P_{21}F^{\rT} G +P_{22} c^{\rT} G^{\rT}G)dJ_2,\\
\label{dVde}
    \d_e V
    & =
        \Gamma_{21} C^{\rT} + Q_{21} BD^{\rT} + Q_{22}e
        -\bA(\Gamma_{22}\Theta_2^{-1})e \wt{J}_1.
\end{align}
Here, $\bA(N):= \frac{1}{2}(N-N^\rT)$ is the antisymmetrizer of matrices, the matrix $\wt{J}_1$ is given by (\ref{tJ1}), and the Gramians $P:= (P_{jk})_{1\< j,k\< 2}$, $Q:= (Q_{jk})_{1\< j,k\< 2}$ and the Hankelian $\Gamma:= (\Gamma_{jk})_{1\< j,k\< 2}$ of the closed-loop system in (\ref{Plim}), (\ref{Qlim}), (\ref{Hank}) are partitioned into square blocks of order $n$.
\hfill$\blacksquare$
\end{thm}
%%%%%%%%%%%%%%%%%%%%%%%%%%%%%%%%%%%%%%%%%%%%%%%%%%%%%%%%%%%%%%%%%%%%%%%%%%%%%%%%%%%%%%%%%%%%%%%%%%%%%%%%%%%%%%%

The first-order necessary conditions of optimality for the CQLQG control problem (\ref{CQLQG}) are obtained by equating to zero the Frechet derivative (\ref{dVdPi}), so that $\d_\Pi V=0$ yields a set of nonlinear matrix algebraic equations consisting of $\d_{R_2}V=0$, $\d_bV = 0$, $\d_e V=0$ from (\ref{dVdR})--(\ref{dVde}) in combination with (\ref{PALE}), (\ref{QALE}), (\ref{Hank}) which capture the dynamics of the closed-loop system. These equations for a locally optimal coherent quantum controller can be solved numerically by using the gradient descent \cite{SVP_2017}, which requires a stabilizing controller as an initial approximation. The presence of the coherent quantum stabilization problem as part of the gradient descent solution of the CQLQG control problem motivates the following alternative approach which employs
a relaxed version of the stability constraint.

\section{Discounted CQLQG control problem}
\label{sec:CQLQGT}

While the quadratic cost functional $V$ in (\ref{V}) and its representations (\ref{VP}), (\ref{VQH}) are well-defined for Hurwitz matrices $\cA$, the following discounted version \cite{B_1965} of the mean square  performance criterion  applies to a wider class of coherent quantum controllers, where the matrix  $\cA$ is not necessarily Hurwitz.
To this end, for a given $T>0$, consider a linear functional $\sE_{T}$ which maps a quantum process $\varphi$ (which can also be a deterministic matrix-valued function of time) to the weighted time average
\begin{equation}
\label{sET}
    \sE_{T}\varphi
    :=
    \tfrac{1}{T}
    \int_{\mR_+}
    \re^{-t/T}
    \bE \varphi(t)
    \rd t.
\end{equation}
Here, the weighting function  $\frac{1}{T} \re^{-t/T}$  is the density of the exponential probability distribution \index{exponential distribution} with mean value $T$ which plays the role of an effective time horizon (ETH) for averaging $\bE \varphi(t)$ over time $t\>0$.  In the framework of such averaging,  the relative importance of the quantity of interest decays exponentially, with $T$ specifying the decay time scale.
Therefore, the time average  in (\ref{sET}) is organised as the discounted cost functionals \index{discounted cost functional} in dynamic programming problems \cite{B_1965}. In particular, if $\bE \varphi(t)$   is right-continuous at $t=0$, then  $\lim_{T\to 0+} \sE_{T} \varphi = \bE \varphi(0)$. Moreover, if $\bE \varphi(t)$ is an analytic function of time $t\>0$, then, by using the moments $\int_{\mR_+}t^k \re^{-t}\rd t = k!$ of the standard exponential  distribution (or the Euler Gamma function at positive integers), it follows from (\ref{sET}) that \begin{align}
\nonumber
    \sE_{T}\varphi
    & =
    \sum_{k=0}^{+\infty}
    \tfrac{1}{k!}
    \d_t^k
    \bE \varphi(t)\big|_{t=0}
    \tfrac{1}{T}
    \int_{\mR_+}
    t^k
    \re^{-t/T}
    \rd t\\
\nonumber
    & =
    \sum_{k=0}^{+\infty}
        T^k
    \d_t^k\bE \varphi(t)\big|_{t=0}\\
\label{sET1}
    & =
    (\cI - T \d_t)^{-1}\bE \varphi(t)\big|_{t=0},
\end{align}
where $\cI$ is the identity operator.
This series is absolutely convergent for sufficiently small values of the ETH $T$ in the sense that
  $T <
  \frac{1}{\limsup_{k\to +\infty}
  \sqrt[k]{|\d_t^k\bE \varphi(t)|_{t=0}|}}$,
where the radius of convergence is computed through the Cauchy-Hadamard theorem. The differential operator on the right-hand side of (\ref{sET1}) can  also be obtained formally as
$
    \tfrac{1}{T}\int_{\mR_+} \re^{-v/T} \re^{v\d_t}\rd v
    =
    \int_{\mR_+} \re^{-v(\cI-T\d_t)}\rd v = (\cI-T\d_t)^{-1}
$
in view of the representation $f(v) = \re^{v\d_t}f(t)|_{t=0}$.
At the other extreme, the infinite-horizon average of $\varphi$ is defined by
\begin{equation}
\label{sEinf}
    \sE_{\infty} \varphi
    :=
    \lim_{T\to +\infty} \sE_{T}\varphi
    =
    \lim_{T\to +\infty}
    \Big(
        \tfrac{1}{T}
        \int_0^T
        \bE \varphi(t)
        \rd t
    \Big),
\end{equation}
provided these limits exist. The second equality in (\ref{sEinf}), whose right-hand side is the Cesaro mean of $\bE \varphi$, follows from the integral version of the Hardy-Littlewood Tauberian theorem \cite{F_1971} since the integral in (\ref{sET}) is the Laplace transform (evaluated at $s=\frac{1}{T}$) of the function $t\mapsto\bE \varphi(t)$.
Therefore, the quantity $V$ in (\ref{V}) can be represented as  the limit
$    V
    =
    \tfrac{1}{2}
    \sE_{\infty}
    (\cZ^{\rT}\cZ)
    =
    \lim_{T\to +\infty}
    V_T
$
of the discounted averages of the process $\cZ^{\rT}\cZ$ in (\ref{cZcZ}):
\begin{equation}
\label{VT}
      V_T
    :=
    \tfrac{1}{2}
    \sE_T
    (\cZ^{\rT}\cZ)
    =
    \tfrac{1}{2T}
    \int_{\mR_+}
    \re^{-t/T}
    \bE (\cZ(t)^{\rT}\cZ(t))
    \rd t.
\end{equation}
In comparison with the infinite-horizon mean square cost functional $V$, its discounted version $V_T$ is applicable to arbitrary (not necessarily stabilizing) coherent quantum controllers for all sufficiently small values of $T$.

%%%%%%%%%%%%%%%%%%%%%%%%%%%%%%%%%%%%%%%%%%%%%%%%%%%%%%%%%%%%%%%%%%%%%%%%%%%%%%%%%%%%%%%%%%%%%%%%%%%
\begin{thm}
\label{th:VT}
Suppose the closed-loop system (\ref{closed_ZX}) is driven by vacuum input fields and satisfies (\ref{cX0good}).
Also, let the ETH $T>0$ satisfy
\begin{equation}
\label{Tmax}
    T < \tfrac{1}{2\max(0,\, \ln \br(\re^\cA))},
\end{equation}
where $\br(\cdot)$ is the spectral radius.
Then the discounted mean square cost $V_T$ in (\ref{VT}) can be computed as
\begin{equation}
\label{VPT}
  V_T
  =
  \tfrac{1}{2}
  \bra
    \cC^{\rT}\cC,
    P_T
  \ket
  =
  \tfrac{1}{2}
  \bra
        Q_T,
        \tfrac{1}{T}\Sigma + \cB\cB^\rT
  \ket
  =
    -
    \bra
        \cA_T,
        \Gamma_T
    \ket.
\end{equation}
Here, the matrix
\begin{equation}
\label{PT}
  P_T:= \Re \sE_{T}(\cX\cX^\rT)
\end{equation}
of the real parts of the discounted second moments of the system variables in (\ref{xx}) is a unique solution of the ALE
\begin{equation}
\label{PTALE}
    \cA_{T}
    P_T
    +
    P_T
    \cA_{T}^{\rT} + \tfrac{1}{T}\Sigma  + \cB\cB^{\rT}= 0
\end{equation}
with the Hurwitz matrix
\begin{equation}
\label{AT}
    \cA_T:= \cA - \tfrac{1}{2T}I_{2n}
\end{equation}
and the matrix  $\Sigma $ given by (\ref{P0}). Also, $Q_T$ is the observability Gramian of the pair $(\cA_T, \cC)$ satisfying the ALE
\begin{equation}
\label{QTALE}
    \cA_T^{\rT}
    Q_T
    +
    Q_T
    \cA_T + \cC^{\rT}\cC= 0,
\end{equation}
and
\begin{equation}
\label{HankT}
  \Gamma_T:= Q_TP_T
\end{equation}
is the corresponding Hankelian.
\hfill$\square$
\end{thm}
%%%%%%%%%%%%%%%%%%%%%%%%%%%%%%%%%%%%%%%%%%%%%%%%%%%%%%%%%%%%%%%%%%%%%%%%%%%%%%%%%%%%%%%%%%%%%%%%%%%
\begin{pf}
The first equality in (\ref{VPT}) is obtained by substituting (\ref{cZcZ}) into (\ref{VT}). In view of (\ref{Xi}), (\ref{Sigma1}),  the matrix $P_T$ in (\ref{PT}) can be computed as
\begin{align}
\nonumber
    P_T & =
    \tfrac{1}{T}
    \int_{\mR_+}
    \re^{-t/T}
    \Xi(t)
    \rd t\\
\nonumber
    & =
    \tfrac{1}{T}
    \int_{\mR_+}
    \re^{t\cA_T}
    \Sigma
    \re^{t\cA_T^{\rT}}
    \rd t
    +
    \int_{\mR_+}
    \Big(
        \tfrac{1}{T}
    \int_{\tau}^{+\infty}
    \re^{-(t-\tau)/T}
    \rd t
    \Big)
        \re^{\tau \cA_T}
        \cB\cB^{\rT}
        \re^{\tau \cA_T^{\rT}}
        \rd \tau
    \\
\label{PTALE1}
    & =
    \int_{\mR_+}
    \re^{t\cA_T}
    \big(
    \tfrac{1}{T}\Sigma  + \cB\cB^{\rT}
    \big)
    \re^{t\cA_T^{\rT}}
    \rd t,
\end{align}
from where the ALE (\ref{PTALE}) follows. Alternatively, (\ref{PTALE}) can be established by applying the Laplace transform to the Lyapunov ODE (\ref{LODE}) and using the property that $TP_T$ is the Laplace transform $\int_{\mR_+}\re^{-st}\Xi(t)\rd t$ of $\Xi$ at $s=\frac{1}{T}$, so that
$
    P_T - \Sigma  = T(\cA P_T +P_T\cA^\rT + \cB\cB^\rT)
$, which indeed leads to (\ref{PTALE}).
 However, (\ref{PTALE1}) explicitly  shows that
the integrals are convergent since
the matrix $\cA_{T}$ in (\ref{AT}) is Hurwitz due to (\ref{Tmax}). Similarly to (\ref{VQH}), the second equality in (\ref{VPT}) follows from the first one in combination with (\ref{QTALE}), (\ref{PTALE}) as
\begin{align}
\nonumber
  V_T
  & =
  -
  \tfrac{1}{2}
  \bra
    \cA_T^\rT Q_T +Q_T \cA_T,
    P_T
  \ket\\
\nonumber
      &
      =
      -
  \tfrac{1}{2}
  \bra
     Q_T,
    \cA_T P_T + P_T \cA_T^\rT
  \ket\\
\label{VTQH}
  & =
  \tfrac{1}{2}
  \bra
    Q_T, \tfrac{1}{T}\Sigma  + \cB\cB^\rT
  \ket.
\end{align}
The last equality in (\ref{VPT}) is obtained from the first equality in (\ref{VTQH}) in view of (\ref{HankT}).
\hfill$\blacksquare$
\end{pf}
%%%%%%%%%%%%%%%%%%%%%%%%%%%%%%%%%%%%%%%%%%%%%%%%%%%%%%%%%%%%%%%%%%%%%%%%%%%%%%%%%%%%%%%%%%%%%%%%%%%

The representation of $V_T$ in (\ref{VPT}) is similar to that of $V$ in (\ref{VP}), (\ref{VQH}). However, in contrast to the infinite-horizon mean square cost  $V$, its discounted counterpart $V_T$ depends (affinely) on the initial covariance condition $\Sigma $ from (\ref{P0}) which enters the ALE (\ref{PTALE}). Furthermore, the matrix $P_T$ also satisfies the Heisenberg uncertainty relation, similar to  (\ref{PTheta}), with the property being inherited from $\Sigma $ since
$
    P_T+i\Theta
    =
    \int_0^{+\infty}
    \re^{t\cA_T}
    \big(
        \tfrac{1}{T}
        (\Sigma  + i\Theta)
        +
        \cB\Omega \cB^{\rT}
    \big)
    \re^{t\cA_T^{\rT}}
    \rd t
    \succcurlyeq 0
$
in view of (\ref{PTALE1}). In particular, $P_T+i\Theta\succ 0$ holds if and only if the matrix pair $(\cA_T, {\small\begin{bmatrix}\tfrac{1}{\sqrt{T}} \sqrt{\Sigma +i\Theta} & & \cB\sqrt{\Omega}\end{bmatrix}})$ is controllable. The latter suggests an interpretation of the term $\frac{1}{T}\Sigma  + \cB\cB^{\rT}$ in the ALE (\ref{PTALE}) as that obtained by appropriately augmenting  the matrix $\cB$, which corresponds to additional input fields  used in \cite{VP_2017}.

Since the condition (\ref{Tmax}) (which is equivalent to the matrix $\cA_T$ in (\ref{AT}) being Hurwitz) does not require the matrix $\cA$ itself to be Hurwitz, the discounted mean square cost $V_T$  is applicable to a wider class of coherent quantum controllers whose matrix triples (\ref{Pi}) form the set
\begin{equation}
\label{fPT}
  \fP_T
  :=
  \{
    \Pi \in \mU:\
    \cA_T\ {\rm in}\ (\ref{AT})\ {\rm is\ Hurwitz}
  \}.
\end{equation}
Such controllers will be referred to as \emph{$T$-stabilizing controllers}.
The set $\fP_T$ in (\ref{fPT}) is an open subset of $\mU$ which is monotonically decreasing with respect to $T$ in the sense that $\fP_{T_1}\supset \fP_{T_2}$ for all $0<T_1 < T_2$. Since for any given controller, the matrix $\cA_T$ in (\ref{AT}) is Hurwitz (that is, the controller is $T$-stabilizing)  for all sufficiently small $T>0$, then
\begin{equation}
\label{fP0}
    \lim_{T\to 0+} \fP_T = \bigcup_{T>0}\fP_T = \mU.
\end{equation}
Therefore, smaller values of the ETH $T$ lead to a relaxation  of the internal stability requirement for the controller. At the other extreme (of large $T$), the set $\lim_{T\to +\infty} \fP_T=\bigcap_{T>0}\fP_T$ consists of all those coherent quantum controller parameters $\Pi \in \mU$ which make the closed-loop system at least marginally stable, with $\br(\re^{\cA})\<1$. In this sense, the inclusion $\fP \subset \bigcap_{T>0}\fP_T$ is ``close'' to being an equality.

The above  considerations allow the infinite-horizon CQLQG control problem (\ref{CQLQG}) to be approached as the ``limiting case'' of its discounted  counterpart
\begin{equation}
\label{CQLQGT}
  V_T\to \inf,
  \qquad
  \Pi \in \fP_T,
\end{equation}
as $T\to +\infty$, where the minimization of $V_T$, computed in Theorem~\ref{th:VT}, is over the set (\ref{fPT}) of $T$-stabilizing coherent quantum controllers.

Since the matrix $\cA_T$ in (\ref{AT}) differs from $\cA$ by an additive term, which does not depend on the controller parameters, and the initial covariance $\Sigma $ is also independent of $\Pi$ from (\ref{Pi}), the proof of the following theorem is similar to that of Theorem~\ref{th:dVdRbe}.

%%%%%%%%%%%%%%%%%%%%%%%%%%%%%%%%%%%%%%%%%%%%%%%%%%%%%%%%%%%%%%%%%%%%%%%%%%%%%%%%%%%%%%%%%%%%%%%%%%%%%%%%%%%%%%%
\begin{thm}
\label{th:dVTdRbe}
For any ETH $T>0$, a $T$-stabilizing coherent quantum controller, described by (\ref{xi_eta}), (\ref{a_c}), (\ref{fPT}), is a stationary point of the discounted mean square cost $V_T$ in (\ref{VT})
if and only if
\begin{align}
\label{dVTdR}
    \d_{R_2} V_T
    & =
    -2\bS(\Theta_2 \Gamma_T^{22}) = 0,\\
\label{dVTdb}
    \d_b V_T
    & =
        Q_T^{21}Ed + Q_T^{22} b-\bA(\Gamma_T^{22}\Theta_2^{-1})bJ_2
    -
    \Theta_2^{-1}
    ((\Gamma_T^{12})^{\rT} E + P_T^{21}F^{\rT} G +P_T^{22} c^{\rT} G^{\rT}G)dJ_2 = 0,\\
\label{dVTde}
    \d_e V_T
    & =
        \Gamma_T^{21} C^{\rT} + Q_T^{21} BD^{\rT} + Q_T^{22}e
        -\bA(\Gamma_T^{22}\Theta_2^{-1})e \wt{J}_1 = 0.
\end{align}
Here,
the Gramians $P_T:= (P_T^{jk})_{1\< j,k\< 2}$, $Q_T:= (Q_T^{jk})_{1\< j,k\< 2}$ and the Hankelian $\Gamma_T:= (\Gamma_T^{jk})_{1\< j,k\< 2}$ of the closed-loop system in (\ref{PTALE}), (\ref{QTALE}), (\ref{HankT}) are partitioned into square blocks
\hfill$\blacksquare$
\end{thm}
%%%%%%%%%%%%%%%%%%%%%%%%%%%%%%%%%%%%%%%%%%%%%%%%%%%%%%%%%%%%%%%%%%%%%%%%%%%%%%%%%%%%%%%%%%%%%%%%%%%%

A combination of (\ref{dVTdR})--(\ref{dVTde})  with (\ref{PTALE}), (\ref{QTALE}), (\ref{HankT})
provides a set of first-order necessary conditions of optimality for the discounted CQLQG control problem (\ref{CQLQGT}) in the class of $T$-stabilizing controllers.
If this problem is considered \emph{individually} for a particular value of the ETH $T>0$,   then it is as complicated as its infinite-horizon counterpart (\ref{CQLQG}) despite the above mentioned relaxation  $\fP_T \supset \fP$ of the internal stability constraint. However, the \emph{family} of these problems can be solved successively by starting from the limit at $T\to 0+$ and evolving the solution as $T$ varies from $0$ to $+\infty$. The smoothness of $V_T(\Pi)$ on the set $\{(T,\Pi): T>0, \Pi \in \fP_T\}$ allows this  solution
to be evolved according to a differential equation. This approach is similar to the homotopy method \cite{MB_1985} for solving sets of nonlinear equations (such as cross-coupled Riccati equations), depending on a scalar parameter, and employs the second-order Frechet derivatives of the cost being minimized.

%%%%%%%%%%%%%%%%%%%%%%%%%%%%%%%%%%%%%%%%%%%%%%%%%%%%%%%%%%%%%%%%%%%%%%%%%%%%%%%%
\section{Second-order optimality conditions and strongly locally optimal controllers}\label{sec:sec}
%%%%%%%%%%%%%%%%%%%%%%%%%%%%%%%%%%%%%%%%%%%%%%%%%%%%%%%%%%%%%%%%%%%%%%%%%%%%%%%%

The cost functional $V_T$ in (\ref{VT}) is invariant under the Lie group of symplectic similarity transformations \cite{VP_2013a,SVP_2017}
\begin{equation}
\label{Rbenew}
    \fS_\sigma:
    \Pi\mapsto (\sigma^{-\rT}R_2\sigma^{-1}, \sigma b,\sigma e)
\end{equation}
of the coherent quantum controller parameters (\ref{Pi}) for any symplectic matrices $\sigma\in \mR^{n\x n}$ (in the sense that $\sigma \Theta_2 \sigma^\rT = \Theta_2$):
\begin{equation}
\label{Vinv}
    V_T(\fS_\sigma(\Pi)) = V_T(\Pi).
\end{equation}
Indeed, the matrices $\cC$, $\Xi$ in (\ref{cABC}), (\ref{Xi}) are transformed as
$
    \cC
    \mapsto
    \cC
    {\small\begin{bmatrix}
      I_n& 0\\
      0 & \sigma^{-1}
    \end{bmatrix}}
$ and
$
    \Xi(t)
    \mapsto
    {\small\begin{bmatrix}
      I_n& 0\\
      0 & \sigma
    \end{bmatrix}}
        \Xi(t)
    {\small\begin{bmatrix}
      I_n& 0\\
      0 & \sigma^\rT
    \end{bmatrix}}
$,
whereby $\cC \Xi(t) \cC^\rT$ remains unchanged and so does $\bra \cC^\rT\cC, \Xi(t)\ket$ for any $t\>0$, thus implying the invariance of $V_T = \frac{1}{2T}\int_{\mR_+} \re^{-t/T} \bra \cC^\rT\cC, \Xi(t)\ket \rd t$ in view of (\ref{VPT}). The \emph{tangent subspace}  $\cT(\Pi)$,  generated by the group (\ref{Rbenew}), and its orthogonal complement $\cN(\Pi)$, which will be referred to as the \emph{normal subspace},   can be represented as
\begin{align}
\label{cT}
    \cT(\Pi)
    := &
    \big\{
        (
            -2
            \bS(R_2g),
            gb,
            ge
        )
        \in
        \mU:
        g \in \Theta_2\mS_n
    \big\},\\
\label{cTort}
    \cN(\Pi)
     := &
     \cT(\Pi)^{\bot}
    =
    \big\{
            (\rho, \beta, \eps)\in \mU:
            2R_2\rho - \beta b^{\rT}-\eps e^{\rT}
            \in
            \Theta_2^{-1}\mA_n
    \big\},
\end{align}
see \cite[Lemma~3]{SVP_2017},
and form an orthogonal decomposition of the Hilbert space $\mU$ in (\ref{Pi}):
\begin{equation}
\label{Usplit}
    \mU = \cT(\Pi)\op \cN(\Pi).
\end{equation}
The symplectic invariance (\ref{Vinv}) of $V_T$ implies that $\cT(\Pi)\subset \d_\Pi V_T^\bot$ or equivalently,
\begin{equation}
\label{ort}
    \d_\Pi V_T \in \cN(\Pi).
\end{equation}
In addition to the first-order necessary conditions of optimality for the discounted CQLQG control problem (\ref{CQLQGT}), provided by Theorem~\ref{th:dVTdRbe},
positive semi-definiteness of the Hessian operator
\begin{equation}
\label{Vdiff2}
    \d_{\Pi}^2 V_T
    :=
    (\d_{\Pi_k}\d_{\Pi_j}V_T)_{1\<j,k\<3}
    =
    {\begin{bmatrix}
        \d_{R_2}^2 V_T    & \d_b\d_{R_2}V_T  &   \d_e\d_{R_2}V_T  \\
        \d_{R_2}\d_b V_T  & \d_b^2 V_T       &  \d_e\d_b V_T\\
        \d_{R_2}\d_e V_T  & \d_b \d_e V_T    &  \d_e^2 V_T
    \end{bmatrix}}
\end{equation}
is a second-order necessary condition
for the stationary point to be a local minimum of $V_T$.
The entries of $\d_{\Pi}^2 V_T$ are the second-order partial Frechet derivatives of  $V_T$ with respect to $\Pi_1:= R_2$, $\Pi_2:= b$, $\Pi_3:= e$ in the matrix triple $\Pi:= (\Pi_k)_{1\<k\<3}$ from (\ref{Pi}).
More precisely, $\d_{\Pi_k}\d_{\Pi_j}V_T= \d_{\Pi_j}\d_{\Pi_k}V_T^{\dagger}$ in (\ref{Vdiff2}) are linear operators on appropriate matrix spaces. For example, $\d_{R_2}^2V_T$ is a self-adjoint operator on $\mS_n$, while $\d_e \d_b V_T: \mR^{n\x p_1}\to \mR^{n\x m_2}$ and $\d_b \d_{R_2} V_T: \mR^{n\x m_2}\to \mS_n$. Note that $\d_{\Pi}^2V_T$ is a self-adjoint operator acting on the Hilbert space $\mU$ in (\ref{Pi}) as
\begin{equation}
\label{VHessmat}
    \gamma
    := (\gamma_k)_{1\<k\<3}
    \mapsto
    \d_{\Pi}^2V_T(\gamma)
    :=
    \Big(
        \sum_{k=1}^3
        \d_{\Pi_k}\d_{\Pi_j}V_T(\gamma_k)
    \Big)_{1\< j\< 3}.
\end{equation}
The
positive semi-definiteness condition $\d_{\Pi}^2 V_T\succcurlyeq 0$ is understood in the sense of nonnegativeness of the second variation
\begin{equation}
\label{Vvar2}
    \delta^2 V_T
     =
    \bra
        \delta \Pi,
        \d_{\Pi}^2V_T(\delta \Pi)
    \ket
    =
    \sum_{j,k=1}^3
    \bra
        \delta \Pi_j,
        \d_{\Pi_k}\d_{\Pi_j}V_T(\delta \Pi_k)
    \ket
\end{equation}
for all $\delta \Pi:= (\delta\Pi_k)_{1\< k\< 3} \in \mU$.
We will now consider an algebraic identity
which follows from the symplectic invariance of the functional $V_T$.

%%%%%%%%%%%%%%%%%%%%%%%%%%%%%%%%%%%%%%%%%%%%%%%%%%%%%%%%%%%%%%%%%%%%%%%%%%%%%%%%%%%%%%%%%%%%%%%%%%%%%%%%%
\begin{lem}
\label{lem:inv_hess}
For any $T$-stabilizing coherent quantum controller in (\ref{fPT}),
which is a stationary point of the discounted mean square cost $V_T$ (that is, $\d_{\Pi}V_T=0$),
the Hessian operator $\d_{\Pi}^2 V_T$ in (\ref{VHessmat}) satisfies
\begin{equation}
\label{Hessort}
    \bra
        \d_{\Pi}^2 V_T(\gamma),
        \psi
    \ket
    =
    0,
    \qquad
    \gamma, \psi \in \cT(\Pi),
\end{equation}
where $\cT(\Pi)$ is the tangent subspace given by (\ref{cT}).
\hfill$\square$
\end{lem}
%%%%%%%%%%%%%%%%%%%%%%%%%%%%%%%%%%%%%%%%%%%%%%%%%%%%%%%%%%%%%%%%%%%%%%%%%%%%%%%%%%%%%%%%%%%%%%%%%%%%%%%%%
\begin{pf}
Consider two one-parameter groups of linear transformations $\re^{\lambda \vartheta}$, $\re^{\mu \varpi}$ on the set $\fP$ in (\ref{fP}), with $\lambda, \mu \in \mR$. Their infinitesimal generators  $\vartheta, \varpi: \fP\to \mU$ are specified by fixed but otherwise arbitrary Hamiltonian matrices  $g,h\in \Theta_2 \mS_n$ and map $\Pi$ to the tangent subspace $\cT(\Pi)$ in (\ref{cT}) as
\begin{equation}
\label{gen1_gen2}
    \vartheta(\Pi)
    := (-2\bS(R_2 g),  gb, ge),
    \qquad
    \varpi(\Pi)
    := (-2\bS(R_2 h),  hb, he).
\end{equation}
These groups are subgroups of the symplectic  transformation group described by  (\ref{Rbenew}) since $\re^{\lambda \vartheta} = \fS_{\re^{\lambda g}}$, $\re^{\mu \varpi}=  \fS_{\re^{\lambda h}}$. Hence,
the symplectic invariance of the discounted mean square cost $V_T$ implies that
\begin{equation}
\label{Vconst}
    V_T(\Pi)
    =
    V_T(\re^{\lambda \vartheta}(\re^{\mu\varpi}(\Pi))).
\end{equation}
By differentiating both sides of (\ref{Vconst}) in $\lambda, \mu\in \mR$ (with $\Pi \in \fP_T$ being fixed), it follows that
\begin{align}
\nonumber
    0 & =
    \d_{\mu}\d_{\lambda}
    V(
        \re^{\lambda \vartheta} (\re^{\mu\varpi}(\Pi))
    )
    \big|_{\lambda=\mu=0}\\
\nonumber
    & =
    \d_{\mu}
    \bra
        \d_{\Pi} V(\re^{\mu\varpi}(\Pi)),
        \vartheta (\re^{\mu\varpi}(\Pi))
    \ket
    \big|_{\mu=0}\\
\label{VVV}
    & =
    \bra
        \d_{\Pi} V,
        \vartheta (\varpi(\Pi))
    \ket
    +
    \bra
        \d_{\Pi}^2 V(\varpi(\Pi)),
        \vartheta(\Pi)
    \ket.
\end{align}
If the $T$-stabilizing coherent quantum controller satisfies $\d_{\Pi}V_T = 0$, then the first term on the right-hand side of (\ref{VVV}) vanishes, which leads to (\ref{Hessort}) in view of arbitrariness of the infinitesimal generators in (\ref{gen1_gen2}).
\hfill$\blacksquare$
\end{pf}
%%%%%%%%%%%%%%%%%%%%%%%%%%%%%%%%%%%%%%%%%%%%%%%%%%%%%%%%%%%%%%%%%%%%%%%%%%%%%%%%%%%%%%%%%%%%%%%%%%%%%%%%%

Lemma~\ref{lem:inv_hess} implies that for any $T$-stabilizing coherent quantum controller, satisfying the first-order optimality conditions (\ref{dVTdR})--(\ref{dVTde}) of  Theorem~\ref{th:dVTdRbe}, the quadratic form $\delta^2V_T$ in (\ref{Vvar2}) vanishes for all elements $\delta \Pi \in \cT(\Pi)$ of the tangent subspace in (\ref{cT}). Therefore, the corresponding Hessian operator $\d_{\Pi}^2V_T$ cannot be positive definite on the whole space $\mU$ (except possibly when $\cT(\Pi)$ is trivial).   Moreover, for any locally optimal controller,
\begin{equation}
\label{kerHess}
    \cT(\Pi)
    \subset
    \ker \d_{\Pi}^2 V_T,
\end{equation}
where $\ker(\cdot)$ is the null space.
 Indeed, the condition $\d_{\Pi}^2 V_T\succcurlyeq 0$ ensures the existence of a positive semi-definite self-adjoint  square root $\sqrt{\d_{\Pi}^2 V_T}$ on the Hilbert space $\mU$. Hence, the relations $\big\|\sqrt{\d_{\Pi}^2 V_T}(\gamma)\big\|^2=\bra \gamma, \d_{\Pi}^2 V_T(\gamma)\ket =0$ for all $\gamma \in \cT(\Pi)$ in view of (\ref{Hessort}) imply that $\cT(\Pi)\subset \ker\sqrt{\d_{\Pi}^2 V_T} = \ker\d_{\Pi}^2 V_T$, which establishes (\ref{kerHess}).  This is a manifestation of the fact that every coherent quantum controller has a class of equivalent state-space representations which are related to each other by symplectic similarity transformations (\ref{Rbenew}). Now, if the inclusion (\ref{kerHess}) for a locally optimal coherent quantum controller holds as an equality, that is,
\begin{equation}
\label{kereqHess}
    \cT(\Pi)
    =
    \ker \d_{\Pi}^2 V_T,
\end{equation}
then, in view of (\ref{Usplit}), the decompositions $\mU = \cT(\Pi)\op \cN(\Pi) =  \ker \d_{\Pi}^2 V_T\op \ker \d_{\Pi}^2 V_T^{\bot}$ imply that $\ker \d_{\Pi}^2 V_T^{\bot} = \cN(\Pi)$. In this case, the Hessian operator $\d_{\Pi}^2 V_T$ maps the normal subspace $\cN(\Pi)$ in (\ref{cTort}) onto itself, with the restriction  $\d_{\Pi}^2 V_T|_{\cN(\Pi)}$  being a positive definite operator. Therefore, the fulfillment of (\ref{kereqHess}) for a locally optimal coherent quantum controller means that the singularity of the Hessian operator comes \emph{only} from the symplectic invariance of $V_T$.

A $T$-stabilizing coherent quantum controller (with $\Pi \in \fP_T$ in (\ref{fPT}))  is said to be a \emph{strong  local minimum} in the discounted CQLQG control problem (\ref{CQLQGT}) if it satisfies the first and second-order necessary conditions of optimality,
and the Hessian operator
in (\ref{Vdiff2}) satisfies (\ref{kereqHess}) (and hence, is positive definite on the normal subspace (\ref{cTort})):
\begin{equation}
\label{SLMT}
    \d_{\Pi}V_T = 0,
    \qquad
    \d_{\Pi}^2V_T \succcurlyeq  0,
    \qquad
    \d_{\Pi}^2V_T\big|_{\cN(\Pi)} \succ  0.
\end{equation}
This definition involves only the first and second-order Frechet derivatives of the discounted mean square cost $V_T$ together with the decomposition (\ref{Usplit}).

%%%%%%%%%%%%%%%%%%%%%%%%%%%%%%%%%%%%%%%%%%%%%%%%%%%%%%%%%%%%%%%%%%%%%%%%%%%%%%%%
\section{A homotopy method for optimal coherent quantum controller synthesis}
\label{sec:homo}
%%%%%%%%%%%%%%%%%%%%%%%%%%%%%%%%%%%%%%%%%%%%%%%%%%%%%%%%%%%%%%%%%%%%%%%%%%%%%%%%

In application of the homotopy method \cite{MB_1985}
to the solution of (\ref{CQLQGT}), we will be concerned with ``regular'' paths of strong local minima $\Pi \in \fP_T$ of the discounted costs $V_T$
in the Hilbert space $\mU$.
A continuously differentiable function $0< T\mapsto \Pi_T \in \fP_T$, taking values in the sets (\ref{fPT}) and having a finite limit
\begin{equation}
\label{Pi0}
    \Pi_0:= \lim_{T\to 0+} \Pi_T \in \mU,
\end{equation}
will be referred to as a \emph{normal solution} of the discounted CQLQG control problem (\ref{CQLQGT}) if, for every $T>0$,  it satisfies (\ref{SLMT}) and
\begin{equation}
\label{norm}
    \d_T \Pi_T \in \cN(\Pi_T).
\end{equation}
The zero-horizon limit $\Pi_0$ in (\ref{Pi0}) is free from stability constraints in accordance with (\ref{fP0}).
The normal solutions can be found through
the integration of a differential equation.

%%%%%%%%%%%%%%%%%%%%%%%%%%%%%%%%%%%%%%%%%%%%%%%%%%%%%%%%%%%%%%%%%%%%%%%%%%%%%%%%%%%%%%%%%%%%%%%%%%%%
\begin{thm}
\label{th:homo}
Every normal solution $0<T\mapsto \Pi_T \in \fP_T$  of the discounted CQLQG control problem (\ref{CQLQGT})
satisfies the ODE
\begin{equation}
\label{homo_ODE}
    \d_T \Pi_T
    =
    -
        \d_{\Pi}^2 V_T
        \big|_{\cN(\Pi_T)}^{-1}
    (\d_T \d_{\Pi} V_T).
\end{equation}
Here, the triple
$    \d_T \d_{\Pi}V_T
    =
    (
        \d_T \d_{R_2}V_T,
        \d_T \d_bV_T,
        \d_T \d_eV_T
    )
$
consists of the matrices
\begin{align}
\label{dVdRdT}
    \d_T\d_{R_2} V_T
    = &
    -2\bS(\Theta_2 \d_T\Gamma_T^{22}),\\
\label{dVdbdT}
    \d_T\d_b V_T
    = &
        \d_TQ_T^{21}Ed + \d_TQ_T^{22} b-\bA(\d_T\Gamma_T^{22}\Theta_2^{-1})bJ_2
        -
    \Theta_2^{-1}
    ((\d_T\Gamma_T^{12})^{\rT} E + \d_TP_T^{21}F^{\rT} G +\d_TP_T^{22} c^{\rT} G^{\rT}G)dJ_2,\\
\label{dVdedT}
    \d_T\d_e V_T
    = &
        \d_T\Gamma_T^{21} C^{\rT} + \d_TQ_T^{21} BD^{\rT} + \d_TQ_T^{22}e
        -\bA(\d_T\Gamma_T^{22}\Theta_2^{-1})e \wt{J}_1,
\end{align}
where
the derivatives $\d_T P_T$, $\d_T Q_T$, $\d_T \Gamma_T$ of the Gramians $P_T$, $Q_T$ in (\ref{PTALE}), (\ref{QTALE}) and the Hankelian  $\Gamma_T$ in (\ref{HankT}) are found from
\begin{align}
\label{dPdT}
    \cA_T\d_T P_T + \d_T P_T \cA_T^{\rT} + \tfrac{1}{T^2}(P_T - \Sigma )
     & = 0,\\
\label{dQdT}
    \cA_T^{\rT}\d_T Q_T + \d_T Q_T \cA_T + \tfrac{1}{T^2}Q_T
     & = 0,\\
\label{dGammadT}
    \d_T \Gamma_T
     & =
    \d_T Q_T P_T + Q_T \d_T P_T.
\end{align}
\hfill$\square$
\end{thm}
%%%%%%%%%%%%%%%%%%%%%%%%%%%%%%%%%%%%%%%%%%%%%%%%%%%%%%%%%%%%%%%%%%%%%%%%%%%%%%%%%%%%%%%%%%%%%%%%%%%%
\begin{pf}
The smoothness of $V_T$ on the set $\bigcup_{T>0}\{T\}\x\fP_T$ and the differentiability of the normal solution $\Pi_T$ in $T$ allow
the left-hand side of the first equality in (\ref{SLMT}) to be differentiated as a composite function of $T>0$:
\begin{equation}
\label{VTdot}
    \d_T \d_{\Pi} V_T  + \d_{\Pi}^2 V_T(\d_T \Pi_T) = 0.
\end{equation}
The property
(\ref{ort}) implies that $\d_T \d_{\Pi}V_T \in \cN(\Pi)$ for any $\Pi\in \fP_T$. Therefore, in view of (\ref{norm}) and the  invertibility of the Hessian operator         $\d_{\Pi}^2 V_T$ on the subspace $\cN(\Pi_T)$ (due to the third relation in (\ref{SLMT})), the equation (\ref{VTdot}) can be uniquely solved for $\d_T \Pi_T$, which leads to (\ref{homo_ODE}).
The relations (\ref{dVdRdT})--(\ref{dVdedT}) are obtained by differentiating (\ref{dVTdR})--(\ref{dVTde}) of Theorem~\ref{th:dVTdRbe} in $T$, and (\ref{dPdT})--(\ref{dGammadT}) are established similarly by using (\ref{PTALE})--(\ref{HankT}).
\hfill$\blacksquare$
\end{pf}
%%%%%%%%%%%%%%%%%%%%%%%%%%%%%%%%%%%%%%%%%%%%%%%%%%%%%%%%%%%%%%%%%%%%%%%%%%%%%%%%%%%%%%%%%%%%%%%%%%%%

The Hessian operator $\d_{\Pi}^2 V_T$ in  (\ref{Vdiff2}), whose inverse is used in (\ref{homo_ODE}), can be  computed with the aid of \cite[Lemmas~5, 8]{VP_2013a} (the resulting expressions are cumbersome and omitted for brevity).

We will now discuss the initial condition (\ref{Pi0}) for the ODE (\ref{homo_ODE}). For what follows, the partitioning of the matrix in (\ref{P0}) into square blocks of order $n$ is denoted by $\Sigma := (\Sigma_{jk})_{1\< j,k\< 2}$.

%%%%%%%%%%%%%%%%%%%%%%%%%%%%%%%%%%%%%%%%%%%%%%%%%%%%%%%%%%%%%%%%%%%%%%%%%%%%%%%%%%%%%%%%%%%%%%%%%%%%
\begin{thm}
\label{th:start}
Suppose $\det \Sigma_{22}\ne 0$ in addition to the assumptions of Theorem~\ref{th:dVTdRbe}. Then for any normal solution $0<T\mapsto \Pi_T \in \fP_T$ of the discounted CQLQG control problem (\ref{CQLQGT}),
the initial condition $\Pi_0 := (R_2^0, b_0, e_0)$ in (\ref{Pi0}) satisfies
\begin{equation}
\label{b0}
    b_0 =  \Theta_2 (c_0^\rT d  + \chi (I_{m_2}-d^\rT d))J_2,
    \qquad
    \chi \in \mR^{n\x m_2},
\end{equation}
where
\begin{equation}
\label{c0}
  c_0 = -(G^\rT G)^{-1} G^\rT F\Sigma_{12}\Sigma_{22}^{-1}
\end{equation}
is the corresponding controller output matrix.
Furthermore, if $c_0$ is of full column rank, then
\begin{equation}
\label{e0}
  e_0 =
  -\Sigma_{21}C^\rT
  -
  (c_0^\rT G^\rT G c)^{-1}
  c_0^\rT G^\rT F (\Sigma_{11}C^\rT + BD^\rT).
\end{equation}
\hfill$\square$
\end{thm}
%%%%%%%%%%%%%%%%%%%%%%%%%%%%%%%%%%%%%%%%%%%%%%%%%%%%%%%%%%%%%%%%%%%%%%%%%%%%%%%%%%%%%%%%%%%%%%%%%%%%
\begin{pf}
In view of the ALEs (\ref{PTALE}), (\ref{QTALE}), the Gramians $P_T$, $Q_T$ admit the truncated Taylor series expansions
\begin{align}
\label{Pasy}
    P_T & = \Sigma  + T P' + O(T^2),\\
\label{Qasy}
    Q_T & = TQ' + \tfrac{1}{2}T^2 Q''+O(T^3),
\end{align}
as $T\to 0+$, where
\begin{align}
\label{P'}
    P' &= \cA \Sigma  + \Sigma \cA^\rT + \cB \cB^\rT,\\
\label{Q'}
    Q'
    & = \cC^\rT \cC,\\
\label{Q''}
    Q''
    & = 2(\cA^\rT Q' + Q'\cA).
\end{align}
By substituting (\ref{Pasy})--(\ref{Q''}) into (\ref{HankT}), it follows that
\begin{equation}
\label{Hankasy}
  \Gamma_T = T \Gamma' + \tfrac{1}{2}T^2 \Gamma''+ O(T^3),
\end{equation}
where
\begin{align}
\label{Hank'}
    \Gamma'
     =&  Q'\Sigma
    =
    \cC^\rT \cC \Sigma ,\\
\label{Hank''}
    \Gamma''
    = &
    2Q'P'+Q''\Sigma
    =
    2(\cC^\rT \cC (2\cA \Sigma  + \Sigma \cA^\rT + \cB \cB^\rT)
    +\cA^\rT \cC^\rT \cC\Sigma) .
\end{align}
The asymptotic relations (\ref{Pasy}), (\ref{Qasy}), (\ref{Hankasy}) hold uniformly with respect to the initial covariances $\Sigma $ and the coherent quantum controller parameters $\Pi \in \mU$ over any bounded sets. Their application to the equalities (\ref{dVTdR})--(\ref{dVTde}) at the normal solution $\Pi_T$
leads to
\begin{align}
\label{dVTdR0}
    \d_{R_2} V_T(\Pi_T)
    &=
    -2T\bS(\Theta_2 \Gamma'_{22}) + O(T^2) = 0,\\
\label{dVTdb0}
    \d_b V_T(\Pi_T)
    & =
    -\Theta_2^{-1}(\Sigma_{21}F^{\rT} G +\Sigma_{22} c_0^{\rT} G^{\rT}G)dJ_2+ O(T)
      = 0,\\
\label{dVTde0}
    \d_e V_T(\Pi_T)
    & =
        T(\Gamma'_{21} C^{\rT}
        +
        Q'_{21} BD^{\rT} +
        Q'_{22}e_0
        -
        \bA(\Gamma'_{22}\Theta_2^{-1})e_0 \wt{J}_1)
        +O(T^2)= 0
\end{align}
as $T\to 0+$. Since the matrix $d$ is of full row rank due to (\ref{ddI}), and the matrices $\Theta_2$,  $J_2$  are nonsingular, (\ref{dVTdb0})  implies that
\begin{equation}
\label{cGG}
    \Sigma_{21}F^{\rT} G +\Sigma_{22} c_0^{\rT} G^{\rT}G=0,
\end{equation}
which leads to (\ref{c0}) in view of (\ref{Grank}) and $\det\Sigma_{22}\ne 0$. The representation (\ref{b0}) describes all those matrices $b$ which produce the matrix $c_0$ according to the second equality in (\ref{a_c}). Here, use is made of the matrix $I_{m_2}-d^\rT d$ of projection onto $\ker d$ in view of (\ref{ddI}) along with the property $J_2^{-1} = -J_2$ of the matrix $J_2$ from (\ref{JJJ}).
By combining (\ref{Hank'}) with (\ref{cGG}), and using the matrix $\cC$ from (\ref{cABC}), it follows that
\begin{align}
\nonumber
    \Gamma'_{22}
    & =
    (\cC^\rT \cC \Sigma)_{22}\\
\nonumber
    & =
    c_0^\rT G^\rT
    {\begin{bmatrix}
    F & Gc_0
    \end{bmatrix}}
    { \begin{bmatrix}
    \Sigma_{12}\\
    \Sigma_{22}
    \end{bmatrix}}    \\
\label{Hank'22}
    & =
    c_0^\rT
    (\Sigma_{21}F^{\rT} G +\Sigma_{22} c_0^{\rT} G^{\rT}G)^\rT
    =0.
\end{align}
This makes the leading term in (\ref{dVTdR0}) vanish. Furthermore, substitution of (\ref{Hank'22}) into (\ref{dVTde0}) yields
\begin{equation}
\label{e01}
\Gamma'_{21} C^{\rT}
        +
        Q'_{21} BD^{\rT} +
        Q'_{22}e_0 = 0.
\end{equation}
The relevant blocks of the matrices (\ref{Q'}), (\ref{Hank'})   take the form
\begin{align}
\label{Q'21}
    Q'_{21}
    &=
    (\cC^\rT \cC)_{21}
    =
    c_0^\rT G^\rT F,\\
\label{Q'22}
    Q'_{22}
    &=
    (\cC^\rT \cC)_{22}
    =
    c_0^\rT G^\rT G c_0,\\
\label{Hank'21}
    \Gamma'_{21}
    &=
    (\cC^\rT \cC\Sigma)_{21}
    =
    c_0^\rT G^\rT (F\Sigma_{11} + G c_0\Sigma_{21}).
\end{align}
In combination with the full column rank of $G$ in (\ref{Grank}), the full column rank of $c_0$ implies that $Q_{22}'\succ 0$ in (\ref{Q'22}). Therefore, substitution of (\ref{Q'21})--(\ref{Hank'21}) into (\ref{e01}) leads to
$    e_0
     = -Q'^{-1}_{22}(\Gamma'_{21} C^{\rT}
        +
        Q'_{21} BD^{\rT})
     = -Q'^{-1}_{22}(c_0^\rT G^\rT (F\Sigma_{11} + G c_0\Sigma_{21}) C^{\rT}
        +
        c_0^\rT G^\rT F BD^{\rT})
   = -\Sigma_{21}C^\rT
  -
  Q'^{-1}_{22}
  c_0^\rT G^\rT F (\Sigma_{11}C^\rT + BD^\rT)
$,
which establishes (\ref{e0}).
\hfill$\blacksquare$
\end{pf}
%%%%%%%%%%%%%%%%%%%%%%%%%%%%%%%%%%%%%%%%%%%%%%%%%%%%%%%%%%%%%%%%%%%%%%%%%%%%%%%%%%%%%%%%%%%%%%%%%%%%

The higher-order terms $TP'$, $\frac{1}{2}T^2Q''$, $\frac{1}{2}T^2\Gamma''$ in (\ref{Pasy}), (\ref{Qasy}), (\ref{Hankasy}), which use (\ref{P'}), (\ref{Q''}), (\ref{Hank''}), depend on the controller energy matrix $R_2$. Its initial value $R_2^0$  can be found by considering further terms in the first-order optimality conditions in the proof of Theorem~\ref{th:start} (these calculations are cumbersome and omitted here for brevity). This completes the initial condition $\Pi_0$ for the numerical integration of the homotopy ODE (\ref{homo_ODE}) over  $T>0$. If the resulting normal solution has a finite limit $\Pi_\infty:= \lim_{T\to +\infty} \Pi_T$, then it describes a marginally stabilizing coherent quantum controller. If this controller is stabilizing (that is, $\Pi_\infty \in \fP$), then it satisfies the first and second-order necessary conditions of optimality  $\d_\Pi V=0$, $\d_\Pi^2\succcurlyeq 0$  for the infinite-horizon CQLQG control problem (\ref{CQLQG}) since $V_T$ converges to $V$ together with all its derivatives as $T\to +\infty$ uniformly over any compact subset of $\fP$.

%%%%%%%%%%%%%%%%%%%%%%%%%%%%%%%%%%%%%%%%%%%%%%%%%%%%%%%%%%%%%%%%%%%%%%%%%%%%%%%
\section{Conclusion}
\label{sec:conc}
%%%%%%%%%%%%%%%%%%%%%%%%%%%%%%%%%%%%%%%%%%%%%%%%%%%%%%%%%%%%%%%%%%%%%%%%%%%%%%%

We have outlined a homotopy method for numerical solution of the coherent quantum LQG control problem for linear quantum plants. In order to relax and then gradually tighten the internal stability requirement, the problem  has been approached as a limit for a family of discounted CQLQG control problems parameterized by the effective  time horizon. A class of strongly locally optimal solutions of these problems has been singled out which satisfy a zero-to-infinite horizon homotopy differential equation involving second-order Frechet derivatives of the discounted mean square cost. In comparison with the gradient descent, this ODE has the advantage that its initialization is free from the stability constraints.


\begin{thebibliography}{99}{%\scriptsize
%%==============================================================================
%\bibitem{A_2000}
%S.L.Adler,
%Derivation of the Lindblad operator structure by use of the Ito stochastic calculus,
%\textit{Phys. Lett. A}, vol. 26, 2000, pp. 58--61.
%==============================================================================

%\bibitem{AM_1979}
%B.D.O.Anderson, and J.B.Moore,
%\emph{Optimal Filtering},
%Prentice Hall, New York, 1979.
%==============================================================================

%\bibitem{AM_1989}
%B.D.O.Anderson, and J.B.Moore,
%\emph{Optimal Control: Linear Quadratic Methods},
%Prentice Hall, London, 1989.
%%==============================================================================
%\bibitem[B1983]{B_1983}
%V.P.Belavkin, On the theory of controlling observable quantum
%systems, \emph{Autom. Rem. Contr.}, vol. 44, no. 2, 1983, pp. 178--188.
%%==============================================================================
%\bibitem{B_2010}
%V.P.Belavkin, Noncommutative dynamics and generalized master equations, \emph{Math. Notes}, vol. 87, no. 5, 2010, pp. 636--653.
%%==============================================================================
%\bibitem[BV1985]{BV_1985}
%A.Bensoussan, and J.H.van Schuppen, Optimal control of
%partially observable stochastic systems with an
%exponential-of-integral performance index, \textit{SIAM J. Control
%Optim.}, vol. 23, 1985, pp. 599--613.
%\bibitem{BH_1998}
%D.S.Bernstein, and W.M.Haddad,
%LQG control with an
%$H^{\infty}$ performance bound: a Riccati equation approach,
%\textit{IEEE Trans.
%Automat. Contr.}, vol. 34, no. 3, 1989, pp. 293--305.
%%==============================================================================

\bibitem[B1965]{B_1965}
    D.Blackwell, Discounted dynamic programming, \emph{Ann. Math. Statist.},
    vol. 36, no. 1, 1965, pp. 226--235.
%%==============================================================================
%\bibitem{B_1968}
%P.Billingsley, \emph{Convergence of Probability Measures}, John Wiley \& Sons, New York, 1968.
%%==============================================================================
%\bibitem[B1996]{B_1996}
%A.Boukas, Stochastic control of operator-valued processes in boson Fock space,
%\textit{Russian J. Mathem. Phys.}, vol. 4, no. 2, 1996, pp. 139--150.

%%==============================================================================

%\bibitem[BH2006]{BH_2006}
%L.Bouten, and R.van Handel, On the separation principle of quantum control,
%arXiv:math-ph/0511021v2, August 22, 2006.
%%%==============================================================================
%\bibitem[BVJ2007]{BVJ_2007}
%L.Bouten, R.Van Handel, M.R.James,
%An introduction to quantum filtering,
%\emph{SIAM J. Control Optim.}, vol. 46, no. 6, 2007, pp. 2199--2241.


%%==============================================================================
%\bibitem{B_1986}
%W.M.Boothby,
%\textit{An Introduction to Differentiable Manifolds and Riemannian Geometry}, 2nd Ed.,
%Academic Press, London, 1986.

%\bibitem{CT_2006}
%T.M.Cover, and J.A.Thomas, \textit{Elements of Information Theory},
%Wiley, New York, 2006.

%%==============================================================================
%\bibitem{CG_2010}
%D.J.Cross, and R.Gilmore,
%A Schwinger disentangling theorem,
%\emph{J. Math. Phys.}, vol. 51, 2010, pp. 103515-1--5.
%%%==============================================================================
%\bibitem{CH_1971}
%C.D.Cushen, and R.L.Hudson, A quantum-mechanical central limit theorem,
%\emph{J. Appl. Prob.}, vol. 8, no. 3, 1971, pp. 454--469.
%%%==============================================================================
%\bibitem[DDJW2006]{DDJW_2006}
%C.D'Helon, A.C.Doherty, M.R.James, and S.D.Wilson,
%Quantum risk-sensitive control,
%Proc. 45th IEEE CDC,
%San Diego, CA, USA, December 13--15, 2006, pp. 3132--3137.


%%%==============================================================================
%\bibitem{D_1970}
%C.Doleans-Dade, Quelques applications de la formule de changement
%de variables pour les semimartingales, \textit{Z. Wahrscheinlichkeitstheorie verw.}, vol. 16, 1970, pp. 181--194.
%%==============================================================================

%\bibitem{DE_1997}
%P.Dupuis, and R.S.Ellis,
%\textit{A Weak Convergence Approach to the Theory of Large Deviations},
%Wiley, New York, 1997.
%%%==============================================================================
%\bibitem[DJP2000]{DJP_2000}
%P.Dupuis, M.R.James, and I.R.Petersen, Robust properties of risk-sensitive control,
% \textit{Math. Control Signals Syst.}, vol.  13, 2000, pp. 318--332.
% %%==============================================================================
%\bibitem[EB2005]{EB_2005}
%S.C.Edwards, and V.P.Belavkin,
%Optimal quantum filtering and
%quantum feedback control,
%arXiv:quant-ph/0506018v2, August 1,  2005.
%%==============================================================================
%\bibitem{E_1998}
%L.C.Evans,
%\textit{Partial Differential Equations},
%American Mathematical Society, Providence, 1998.

%%%==============================================================================
\bibitem[F1971]{F_1971}
W.Feller, \emph{An Introduction to Probability Theory and Its Applications. Vol. II}, 2nd Ed., John Wiley \& Sons, New York, 1971.

%\bibitem[F1989]{F_1989}
%G.B.Folland, \emph{Harmonic Analysis in Phase Space}, Princeton University Press, Princeton, 1989.



%\bibitem[GZ2004]{GZ_2004}
%C.W.Gardiner, and P.Zoller,
%\textit{Quantum Noise}.
%Springer, Berlin, 2004.
%%==============================================================================
%\bibitem{GHP_2009}
%P.Gibilisco, F.Hiai, and D.Petz,
%Quantum covariance, quantum Fisher information, and the uncertainty principle,
%\textit{IEEE Trans.
%Inform. Theory.}, vol. 55, no. 1, 2009, pp. 439--443.
%%%==============================================================================
%\bibitem{GS_2004}
%I.I.Gikhman, and A.V.Skorokhod,
%\textit{The Theory of Stochastic Processes}, Springer, Berlin,
%2004.

%\bibitem{G_1994}
%M.S.Ginovian, On Toeplitz type quadratic functionals of stationary
%Gaussian processes, {\it Probab. Theory Relat. Fields}, vol. 100, 1994,
%pp. 395--406.
%%==============================================================================
%\bibitem{G_1965}
%S.Golden, Lower bounds for the Helmholtz function, \textit{Phys. Rev.}, vol.
%137, 1965, pp. B1127--B1128.
%%%==============================================================================
%\bibitem{GKS_1976}
%V.Gorini, A.Kossakowski, E.C.G.Sudarshan, Completely positive dynamical semigroups of N-level systems,
%\textit{J. Math. Phys.}, vol. 17, no. 5, 1976, pp. 821--825.
%%==============================================================================
\bibitem[GJ2009]{GJ_2009}
J.Gough, and M.R.James,
Quantum feedback networks: Hamiltonian
formulation,
\emph{Commun. Math. Phys.},  vol. 287, 2009, pp. 1109--1132.
%%==============================================================================
%\bibitem{G_2006}
%M.de~Gosson,
%\textit{Symplectic Geometry and Quantum Mechanics},
%Birk\-h\"{a}user, Basel, 2006.
%%==============================================================================
%\bibitem{G_1977}
%H.W.Guggenheimer,
%\textit{Differential Geometry},
%Dover, New York, 1977.
%%%==============================================================================
%\bibitem[H2008]{H_2008}
%N.J.Higham,
%\textit{Functions of Matrices},
%SIAM, Philadelphia,  2008.
%%%%==============================================================================
%\bibitem{H_1991}
%A.S.Holevo, Quantum stochastic calculus,
%\textit{J. Sov. Math.},
%vol. 56, no. 5, 1991, pp. 2609--2624.
%%==============================================================================

%\bibitem{H_1996}
%A.S.Holevo,
%Exponential formulae in quantum stochastic calculus,
%\textit{Proc. Roy. Soc. Edinburgh}, vol. 126A, 1996, pp. 375--389.
%%==============================================================================
\bibitem[H2001]{H_2001}
A.S.Holevo, \textit{Statistical Structure of Quantum Theory}, Springer, Berlin, 2001.
%%==============================================================================
%\bibitem{H_2010}
%R.L.Hudson,
%Quantum Bochner theorems and incompatible observables,
%\emph{Kybernetika}, vol. 46, no. 6, 2010, pp. 1061--1068.
%%==============================================================================
\bibitem[HJ2007]{HJ_2007}
R.A.Horn, and C.R.Johnson,
\textit{Matrix Analysis},
Cambridge
University Press, New York, 2007.
%%==============================================================================
%\bibitem{HM_1994}
%U.Helmke, and J.B.Moore,
%\textit{Optimization and Dynamical Systems},
%Springer, London, 1994.
%%==============================================================================
\bibitem[HP1984]{HP_1984}
R.L.Hudson,  and K.R.Parthasarathy,
Quantum Ito's formula and stochastic evolutions,
\textit{Commun. Math. Phys.}, vol.  93, 1984, pp. 301--323.
%%%==============================================================================
%\bibitem[H2018]{H_2018}
%R.L.Hudson, A short walk in quantum probability,
%\textit{Philos. Trans. R. Soc. A}, vol.  376, 2018, pp. 1--13.

%\bibitem{I_1918}
%L.Isserlis, On a formula for the product-moment coefficient of any order of a normal
%frequency distribution in any number of variables, \emph{Biometrika}, vol. 12, 1918, pp.
%134--139.
%%==============================================================================
%\bibitem{JK_1998}
%K.Jacobs, and P.L.Knight,
%Linear quantum trajectories: applications to continuous projection measurements,
%\textit{Phys. Rev. A}, vol.  57, no. 4, 1998, pp. 2301--2310.
%%%==============================================================================
%\bibitem[J1973]{J_1973}
%D.H.Jacobson, Optimal stochastic linear systems with
%exponential performance criteria and their relation to
%deterministic differential games, \textit{IEEE Trans. Autom.
%Control}, vol. 18, 1973,  pp. 124--31.
%%%==============================================================================
%\bibitem[J2004]{J_2004}
%M.R.James, Risk-sensitive optimal control of quantum systems,
%\emph{Phys. Rev. A},  vol. 69, 2004, pp. 032108-1--14.
%%==============================================================================
%\bibitem[J2005]{J_2005}
%M.R.James, A quantum Langevin formulation of risk-sensitive optimal control,
%\emph{J. Opt. B}, vol. 7, 2005, pp. S198--S207.

%%==============================================================================
\bibitem[JG2010]{JG_2010}
M.R.James, and J.E.Gough, Quantum dissipative systems and feedback control design by interconnection,
\textit{IEEE Trans.
Automat. Contr.}, vol. 55, no. 8, 2010, pp. 1806--1821.
%%==============================================================================
\bibitem[JNP2008]{JNP_2008}
M.R.James, H.I.Nurdin, and I.R.Petersen,
$H^{\infty}$ control of
linear quantum stochastic systems,
\textit{IEEE Trans.
Automat. Contr.}, vol. 53, no. 8, 2008, pp. 1787--1803.
%
%%==============================================================================

%\bibitem{J_1997}
%S.Janson, \emph{Gaussian Hilbert Spaces}, Cambridge University Press, Cambridge, 1997.
%%==============================================================================
%\bibitem[KS1991]{KS_1991}
%I.Karatzas, and S.E.Shreve,
%\emph{Brownian Motion and Stochastic Calculus}, 2nd Ed.,
%Springer, New York, 1991.

%%==============================================================================
%\bibitem{K_1976}
%T.Kato, \textit{Perturbation Theory for Linear Operators}, 2nd Ed., Springer, Berlin, 1976.

%%==============================================================================
%\bibitem{K_1972}
%A.Kossakowski, On quantum statistical mechanics of non-Hamiltonian systems,
%\textit{Rep. Math. Phys.},
%vol. 3, no. 4, 1972, pp. 247--274.
%%==============================================================================

\bibitem[KS1972]{KS_1972}
H.Kwakernaak, and R.Sivan,
\textit{Linear Optimal Control Systems},
Wiley, New York, 1972.
%%%%==============================================================================
%\bibitem{LL_1991}
%L.D.Landau, and E.M.Lifshitz, \textit{Quantum Mechanics: Non-relativistic Theory},
%3rd Ed., Pergamon Press,  Oxford, 1991.
%%==============================================================================
%\bibitem{L_1976}
%G. Lindblad, On the generators of quantum dynamical semigroups,
%\textit{Commun. Math. Phys.}, vol. 48, 1976, pp. 119--130.
%%%==============================================================================
%\bibitem{LS_2001}
%R.S.Liptser, and A.N.Shiryaev,
%\textit{Statistics of Random Processes: Applications}, Springer, Berlin, 2001.
%%%==============================================================================
%\bibitem{L_1968}
%M.Lutzky, Parameter differentiation of exponential operators and the Baker-Campbell-Hausdorff formula,
%\textit{J. Math. Phys.}, vol. 9, no. 7, 1968, pp. 1125--1128.
%%==============================================================================
\bibitem[MP2009]{MP_2009}
A.I.Maalouf, and I.R.Petersen,
Coherent LQG control for a class of linear complex quantum systems,
IEEE European Control Conference, Budapest, Hungary, 23-26 August 2009,
pp. 2271--2276.
%%==============================================================================
%\bibitem{MP_2012}
%A.I.Maalouf, and I.R.Petersen,
%On the physical realizability of a class of nonlinear quantum systems, \textit{submitted}.
%%==============================================================================
%\bibitem{M_2005}
%M.Maggiore,
%\textit{A Modern Introduction to Quantum Field Theory},
%Oxford University Press, New York, 2005.
%%==============================================================================
%\bibitem{M_1954}
%W.Magnus, On the exponential solution of differential equations for a linear operator,
%\textit{Comm. Pure Appl. Math.}, vol. 7, no. 4, 1954, pp. 649--673.
%%==============================================================================
%\bibitem{M_1978}
%J.R.Magnus,
%The moments of products of quadratic forms in normal variables,
%\textit{Statist. Neerland.},  vol. 32, 1978, pp. 201--210.
%%==============================================================================
%\bibitem{M_1988}
%J.R.Magnus,
%\textit{Linear Structures},
%Oxford University Press, New York, 1988.
%
%%==============================================================================
\bibitem[MB1985]{MB_1985}
M.Mariton, and P.Bertrand,
A homotopy algorithm for solving coupled
Riccati equations, \textit{Optim. Contr. Appl. Methods}, vol. 6, 1985, pp. 351--357.


%\bibitem{MP_2009}
%A.Mazurov, and P.Pakshin,
%Stochastic dissipativity with risk-sensitive storage
%function and related control problems, \emph{ICIC Express Letters}, vol. 3, no. 1, 2009, pp. 53--60.

%%==============================================================================

%\bibitem[M1998]{M_1998}
%E.Merzbacher,
%\textit{Quantum Mechanics}, 3rd Ed.,
%Wiley, New York, 1998.
%%%==============================================================================
%\bibitem[M1995]{M_1995}
%P.-A.Meyer,
%\textit{Quantum Probability for Probabilists},
%Springer, Berlin, 1995.
%%==============================================================================
\bibitem[MJ2012]{MJ_2012}
Z.Miao, and M.R.James, Quantum observer for linear quantum
stochastic systems, Proc. 51st IEEE Conf. Decision Control, Maui,
Hawaii, USA, December 10-13, 2012, pp. 1680--1684.

%\bibitem{MG_1990}
%D.Mustafa,  and K.Glover, \emph{Minimum Entropy $H^{\infty}$-Control}, Springer Verlag, Berlin,
%%Lecture Notes in Control and Information Sciences, volume 146,
%1990.
%%%==============================================================================
%\bibitem[NC2000]{NC_2000}
%M.A.Nielsen, and I.L.Chuang,
%\textit{Quantum Computation and Quantum Information},
%Cambridge University Press, Cambridge, 2000.
%%==============================================================================
\bibitem[NJP2009]{NJP_2009}
H.I.Nurdin, M.R.James, and I.R.Petersen,
Coherent quantum LQG
control,
\textit{Automatica}, vol.  45, 2009, pp. 1837--1846.
%

%%%==============================================================================
%\bibitem[N2014]{N_2014}
%H.I.Nurdin,
%Quantum filtering for multiple input multiple output systems driven by arbitrary zero-mean jointly Gaussian input fields, \textit{Russian J. Math. Phys.}, vol. 21, no. 3, 2014,  pp. 386--398.

\bibitem[NY2017]{NY_2017}
H.I.Nurdin, and N.Yamamoto,
\textit{Linear Dynamical Quantum Systems},
Springer, Netherlands, 2017.


%%%==============================================================================
%\bibitem{OP_1993}
%M.Ohya, and D.Petz, \textit{Quantum Entropy and Its Use}, Springer-Verlag, Berlin,
%1993.
%%%==============================================================================
%\bibitem[OW2010]{OW_2010}
%M.Ohya, and N.Watanabe,
%Quantum entropy and its applications to quantum
%communication and statistical physics,
%\textit{Entropy}, vol.  12, 2010, pp. 1194--1245.


%%==============================================================================
\bibitem[P1992]{P_1992}
K.R.Parthasarathy,
\textit{An Introduction to Quantum Stochastic Calculus},
Birk\-h\"{a}user, Basel, 1992.
%%==============================================================================
\bibitem[P2010]{KRP_2010}
K.R.Parthasarathy,
What is a Gaussian state?
\textit{Commun. Stoch. Anal.}, vol. 4, no. 2, 2010, pp. 143--160.
%%==============================================================================

%\bibitem[PS1972]{PS_1972}
%K.R.Parthasarathy, and K.Schmidt,
%\emph{Positive Definite Kernels, Continuous Tensor Products, and Central Limit Theorems of Probability Theory},
%Springer-Verlag, Berlin, 1972.
%%==============================================================================

%\bibitem{PS_2015}
%K.R.Parthasarathy,  and R.Sengupta,
%From particle counting to Gaussian tomography,
% \emph{Inf. Dim. Anal., Quant. Prob. Rel. Topics}, vol. 18, no. 4,  2015, pp. 1550023.
%%%==============================================================================
%\bibitem[P2015]{P_2015}
%K.R.Parthasarathy,
%Quantum stochastic calculus and quantum Gaussian processes,
%\textit{Indian Journal of Pure and Applied Mathematics}, 2015, vol. 46, no. 6, 2015, pp. 781--807.
%%%==============================================================================
%\bibitem[P2006]{P_2006}
%I.R.Petersen, Minimax LQG control, \textit{Int. J. Appl. Math. Comput. Sci.}, vol. 16, no. 3, 2006, pp. 309--323.
%%%==============================================================================
%\bibitem{P_2014}
%I.R.Petersen,
%Guaranteed non-quadratic performance for quantum systems with nonlinear uncertainties,
% \textit{American Control Conference (ACC),
%4-6 June 2014},  arXiv:1402.2086 [quant-ph], 10 February 2014.
%%==============================================================================
\bibitem[P2017]{P_2017}
I.R.Petersen,
Quantum linear systems theory,
\textit{Open Automat. Contr. Syst. J.},
vol. 8, 2017, pp. 67--93.
%Proc. 19th Int. Symp. Math. Theor. Networks Syst., Budapest, Hungary, July 5--9, 2010, pp.  2173--2184.

%\bibitem{P_2010}
%I.R.Petersen,
%Quantum linear systems theory,
%Proc. 19th Int. Symp. Math. Theor. Networks Syst., Budapest, Hungary, July 5--9, 2010, pp.  2173--2184.
%%==============================================================================
%
%\bibitem[PJD2000]{PJD_2000}
%I.R.Petersen, M.R.James, and P.Dupuis, Minimax optimal control of stochastic uncertain
%systems with relative entropy constraints, \textit{IEEE Trans. Automat. Contr.}, vol. 45, 2000,
%pp. 398--412.
%%%==============================================================================
%\bibitem{PUS_2000}
%I.R.Petersen, V.A.Ugrinovskii, and A.V.Savkin, \textit{Robust Control Design Using $\cH^\infty$-
%Methods}, Springer, London, 2000.
%%==============================================================================
%\bibitem{PUJ_2012}
%I.R.Petersen, V.A.Ugrinovskii, and M.R.James,
%Robust stability of uncertain linear
%quantum systems,
%\emph{Phil. Trans. R. Soc. A},  vol. 370, 2012, pp. 5354--5363.

%arXiv:1203.2676v1 [quant-ph], 12 March 2012.

%%==============================================================================

%%==============================================================================
%\bibitem{PBGM_1962}
%L.S.Pontryagin, V.G.Boltyanskii,
%R.V.Gamkrelidze, and E.F. Mishchenko,
%\emph{The Mathematical Theory of Optimal Processes},
%Wiley, New York, 1962.
%==============================================================================
%\bibitem{SIS_2004}
%A.Serafini, F.Illuminati, and S.De Siena,
%Symplectic invariants, entropic measures and correlation of Gaussian states,
%\textit{J. Phys. B: At. Mol. Opt. Phys.}, vol. 37, 2004, pp. L21--L28.
%==============================================================================
%

\bibitem[S1994]{S_1994}
J.J.Sakurai,
\textit{Modern Quantum Mechanics},
 Addison-Wesley, Reading, Mass., 1994.
%==============================================================================

%\bibitem{SP_2009}
%A.J.Shaiju, and I.R.Petersen,
%On the physical realizability of
%general linear quantum stochastic differential equations with
%complex coefficients,
%Proc. Joint 48th IEEE Conf. Decision Control \&
%28th Chinese Control Conf.,
%Shanghai, P.R. China, December 16--18, 2009, pp. 1422--1427.

%%==============================================================================
%\bibitem[SPJ2007]{SPJ_2007}
%A.J.Shaiju, I.R.Petersen, and M.R.James,
%Guaranteed cost LQG control of uncertain linear stochastic quantum systems,
%American Control Conference, New York, 9–13 July 2007, pp. 2118--2123.


%==============================================================================
\bibitem[SP2012]{SP_2012}
A.J.Shaiju, and I.R.Petersen,
A frequency domain condition for the physical
realizability of linear quantum systems,
\emph{IEEE Trans. Automat. Contr.}, vol. 57, no. 8, 2012, pp. 2033--2044.

%--------------------------------------------------------------------
%\bibitem{S_1996}
%A.N.Shiryaev, \emph{Probability}, 2nd Ed., Springer, New York, 1996.

\bibitem[SVP2017]{SVP_2017}
A.Kh.Sichani, I.G.Vladimirov, and I.R.Petersen,
A numerical approach to optimal coherent quantum LQG controller design using gradient descent,
\textit{Automatica},
vol. 85, 2017,
pp. 314--326.
%%==============================================================================
%
%\bibitem{S_2000}
%R.Simon,
%Peres-Horodecki separability criterion for continuous variable systems,
%\textit{Phys. Rev. Lett.},
%vol. 84, no. 12, 2000, pp. 2726--2729.
%%==============================================================================
%\bibitem{S_2005}
%B.Simon, \textit{Trace Ideals and Their Applications}, 2nd Ed.,
%American Mathematical Society, Providence, RI, 2005.

%
%%==============================================================================
%%
%\bibitem{SIG_1998}
%R.E.Skelton, T.Iwasaki, and K.M.Grigoriadis,
%\textit{A Unified Algebraic Approach to Linear Control Design},
%Taylor \& Francis, London, 1998.
%%%==============================================================================
%\bibitem{SW_1997}
%H.J.Sussmann, and J.C.Willems,
%300 years of optimal control: from the brachystochrone to the maximum principle,
%\textit{Control Systems}, vol. 17, no. 3, 1997, pp. 32--44.
%%%==============================================================================

%\bibitem{T_1965}
%C.J.Thompson, Inequality with applications in statistical mechanics,
%\textit{J. Math. Phys.}, vol. 6, 1965, pp. 1812--1813.
%\bibitem{V_2008}
%O.V.Viskov, On the Mehler formula for Hermite polynomials, \emph{Dokl. Math.},  vol. 77, no. 1, 2008, pp. 1--4.
%
%%==============================================================================
%\bibitem{V_1971}
%V.S.Vladimirov,
%\textit{Equations of Mathematical Physics},
%M.Dekker,
%New York, 1971.
%%%==============================================================================
%\bibitem[V2002]{V_2002}
%V.S.Vladimirov.
%\textit{Methods of the Theory of Generalized Functions},
%London: Taylor \& Francis, 2002.

%%%%==============================================================================
%\bibitem[VP2010]{VP_2010}
%I.G.Vladimirov, and I.R.Petersen,
%Minimum relative entropy state transitions in linear stochastic
%systems: the continuous time case,
%19th International Symposium on Mathematical Theory of Networks and Systems (MTNS 2010),
%Budapest, Hungary, July 5--9,  2010, pp.  51--58.
%%%==============================================================================
%\bibitem{VP_2010b}
%I.G.Vladimirov, and I.R.Petersen,
%Hardy-Schatten norms of systems, output energy cumulants and linear quadro-quartic  Gaussian control,
%Proc. 19th Int. Symp. Math. Theor. Networks Syst., Budapest, Hungary, July 5--9,  2010, pp.  2383--2390.
%%==============================================================================
%\bibitem{VP_2011a}
%I.G.Vladimirov, and I.R.Petersen,
%A quasi-separation principle and Newton-like scheme for coherent quantum LQG control, 	
%18th IFAC World Congress, Milan, Italy, 28 August--2 September, 2011, pp. 4721--4727.
%(preprint:  arXiv:1010.3125v2 [quant-ph], 15 April 2011).
%%%==============================================================================
%\bibitem[VP2011]{VP_2011b}
%I.G.Vladimirov, and I.R.Petersen,
%A dynamic programming approach to finite-horizon coherent quantum LQG control, 	
%Australian Control Conference, Melbourne, 10--11 November, 2011, pp. 357--362,
%(preprint: {\tt arXiv:1105.1574v1 [quant-ph], 9 May 2011}).
%%%==============================================================================
%\bibitem{VP_2012a}
%I.G.Vladimirov, and I.R.Petersen,
%Gaussian stochastic  linearization for open quantum systems
%using quadratic approximation of Hamiltonians, 	20th Int. Symp. Math. Theor. Networks Syst., Melbourne, Victoria, July 9--13,  2012 (preprint: arXiv:1202.0946v1 [quant-ph], 5 February 2012).
%%==============================================================================
%\bibitem{VP_2012b}
%I.G.Vladimirov, and I.R.Petersen,
%Risk-sensitive dissipativity of linear quantum stochastic systems
%under Lur'e type perturbations of Hamiltonians, 	Proc. AUCC 2012, Sydney, Australia, 15--16 November 2012, pp. 247--252 (preprint:  	
%arXiv:1205.3566v1 [quant-ph], 16 May 2012).
%==============================================================================
%\bibitem{VP_2012c}
%I.G.Vladimirov, and I.R.Petersen,
%Characterization and moment stability analysis of quasilinear quantum stochastic systems with quadratic coupling to external fields, Proc. 51st Conference on Decision and Control, IEEE, Maui, Hawaii, USA, 10-13 December  2012, pp. 1691--1696.

%%==============================================================================
\bibitem[VP2013a]{VP_2013a}
I.G.Vladimirov, and I.R.Petersen,
A quasi-separation principle and Newton-like scheme for coherent quantum LQG control,
\emph{Syst. Contr. Lett.}, vol. 62, no. 7, 2013, pp. 550--559.
%%==============================================================================
\bibitem[VP2013b]{VP_2013b}
I.G.Vladimirov, and I.R.Petersen,
Coherent quantum filtering for physically realizable linear quantum plants,
Proc. European Control Conference, IEEE, Zurich, Switzerland, 17-19 July 2013,  pp. 2717--2723.
%%==============================================================================

%\bibitem{V_2015c}
%I.G.Vladimirov,
%Evolution of quasi-characteristic functions in quantum stochastic systems with Weyl quantization
%of energy operators, {\tt arXiv:1512.08751 [math-ph], 29 December 2015}.
%%==============================================================================


%%==============================================================================

%\bibitem[VPJ2018a]{VPJ_2018a}
%I.G.Vladimirov, I.R.Petersen, and M.R.James, Multi-point Gaussian states, quadratic–exponential cost functionals, and large deviations estimates for linear quantum stochastic systems, \textit{Appl. Math. Optim.}, 2018, pp. 1--55.

%\bibitem[VPJ2018b]{VPJ_2018b}
%I.G.Vladimirov, I.R.Petersen, and M.R.James,
%Risk-sensitive performance criteria and robustness of quantum systems with a relative entropy description of state uncertainty, 23rd International Symposium on Mathematical Theory of Networks and Systems (MTNS 2018),
%Hong Kong University of Science and Technology, Hong Kong, July 16-20, 2018, pp. 482--488.

% \bibitem[VPJ2018c]{VPJ_2018c}
%I.G.Vladimirov, I.R.Petersen, and M.R.James, Parametric randomization,  complex symplectic factorizations, and quadratic-exponential functionals for Gaussian quantum states, \textit{Inf.-Dim. Anal., Quant. Prob. Rel. Topics}, accepted
% 	(preprint {\tt arXiv:1809.06842 [quant-ph], 18 September 2018}).



%\bibitem[VPJ2019a]{VPJ_2019a}
%I.G.Vladimirov, I.R.Petersen, and M.R.James, Lie-algebraic connections between two classes of risk-sensitive performance criteria for linear quantum stochastic systems, SIAM Conference on Control and Its Applications (CT19), June 19-21, 2019, Chengdu, China, pp. 30--37 (preprint: {\tt  	arXiv:1903.00710 [math-ph], 2 March 2019}).

%\bibitem[VPJ2019b]{VPJ_2019b}
%I.G.Vladimirov, I.R.Petersen, and M.R.James,
%A quantum Karhunen-Loeve expansion and quadratic-exponential functionals for linear quantum stochastic systems, 58th Conference on Decision and Control (CDC2019), Nice, France, 11-13 December 2019, accepted (preprint
%{\tt  	arXiv:1904.03265 [math.PR], 5 April 2019}).

%\bibitem[VJP2019]{VJP_2019}
%I.G.Vladimirov, M.R.James, and I.R.Petersen,
%A Karhunen-Loeve expansion for one-mode open quantum harmonic
%oscillators using the eigenbasis of the two-point commutator kernel, 2019 Australian and  New Zealand Control Conference (ANZCC2019),
%Auckland, New Zealand, 27-29 November 2019, accepted (preprint
%{\tt  	arXiv:1909.07377 [quant-ph], 16 September 2019}).


%\bibitem[VPJ2019c]{VPJ_2019c}
%I.G.Vladimirov, I.R.Petersen, and M.R.James,
%A Girsanov type representation of quadratic-exponential cost
%functionals for linear quantum stochastic systems, submitted to ECC2020
%(preprint
%{\tt  	arXiv:1911.01539 [quant-ph], 4 November 2019}).

%\bibitem[VPJ2019d]{VPJ_2019d}
%I.G.Vladimirov, I.R.Petersen, and M.R.James,
%Frequency-domain computation of
%quadratic-exponential cost functionals for linear
%quantum stochastic systems, submitted to IFAC2020 (preprint
%{\tt  	arXiv:1911.03031v1 [quant-ph] 8 November 2019}).



%I.G.Vladimirov, I.R.Petersen, and M.R.James
%Multi-point Gaussian states, quadratic-exponential cost functionals, and large deviations estimates for linear quantum stochastic systems, {\tt arXiv:1707.09302 [math.OC], 28 July 2017}.

\bibitem[VP2018]{VP_2018}
I.G.Vladimirov, and I.R.Petersen,
Direct coupling coherent quantum observers with discounted mean square performance criteria and penalized back-action,  	
(preprint:  arXiv:1809.00389 [cs.SY], 2 September 2018).



\bibitem[VP2017]{VP_2017}
S.L.Vuglar, and I.R.Petersen,
Quantum noises, physical realizability and coherent quantum feedback control,
\textit{IEEE Trans. Automat. Contr.}, vol. 62, no. 2, 2017, pp. 998--1003.

%%==============================================================================

%\bibitem{W_1972}
%J.C.Willems, Dissipative dynamical systems.
%Part I: general theory, Part II: linear systems with quadratic supply rates, \emph{Arch. Rat. Mech. Anal.}, vol. 45, no. 5, 1972, pp. 321--351, 352--393.
%%==============================================================================
%\bibitem{W_1936}
%J.Williamson,
%On the algebraic problem concerning the normal forms of linear dynamical systems,
%\textit{Am. J. Math.}, vol. 58, no. 1, 1936, pp. 141--163.
%%==============================================================================
%\bibitem{W_1937}
%J.Williamson,
%On the normal forms of linear canonical transformations in dynamics,
%\textit{Am. J. Math.}, vol. 59, no. 3, 1937, pp. 599--617.
%%%%==============================================================================
%\bibitem[WM2008]{WM_2008}
%D.F.Walls, and G.J.Milburn,
%\emph{Quantum Optics}, 2nd Ed., Springer, Berlin, 2008.
%%%==============================================================================
%\bibitem[W1981]{W_1981}
%P.Whittle, Risk-sensitive linear/quadratic/Gaussian
%control, \textit{Adv. Appl. Probab.}, vol. 13,  1981, pp. 764--77.
%%==============================================================================
%\bibitem{W_1967}
%R.M.Wilcox,
%Exponential operators and parameter differentiation in quantum physics,
%\emph{J. Math. Phys.}, vol.  8, no. 4, 1967, pp. 962--982.
%%%==============================================================================
%\bibitem[WM2010]{WM_2010}
%H.M.Wiseman, and G.J.Milburn,
%\emph{Quantum measurement and control},
%Cambridge University Press,
%Cambridge, 2010.
%%%==============================================================================
%\bibitem[YB2009]{YB_2009}
%N.Yamamoto, and L.Bouten,
%Quantum risk-sensitive estimation and robustness,
%\emph{IEEE Trans. Automat. Contr.}, vol. 54, no. 1, 2009, pp. 92--107.
%%==============================================================================
%\bibitem{Y_2009}
%M.Yanagisawa, Non-Gaussian state generation from linear elements via feedback,
%\textit{Phys. Rev. Lett.}, vol. 103, no. 20, pp. 203601-1--4.
%%==============================================================================

%\bibitem{Y_1980}
%K.Yosida, \emph{Functional Analysis}, 6th Ed., Springer, Berlin, 1980.
%%==============================================================================
\bibitem[ZJ2011]{ZJ_2011a}
G.Zhang, and M.R.James,  Direct and indirect couplings in coherent feedback control of linear quantum systems,
\textit{IEEE Trans. Automat. Contr.}, vol. 56, no. 7, 2011, 1535--1550.
%%==============================================================================
%\bibitem{ZJ_2011b}
%G.Zhang, and M.R.James, On the response of linear quantum stochastic systems to single-photon inputs and pulse shaping of photon wave packets, Proc. Australian Control Conference, Melbourne, 10--11 November, 2011, pp. 62--67.
%%%==============================================================================
%\bibitem[ZJ2012]{ZJ_2012}
%G.Zhang, and M.R.James,
%Quantum feedback networks and control: a brief survey,
%\emph{Chinese Sci. Bull.}, vol. 57, no. 18, 2012, pp. 2200--2214,
%arXiv:1201.6020v2 [quant-ph], 26 February 2012.
%%
}
\end{thebibliography}
\end{document}